\documentclass[10pt,amsmath,amssymb,amsfont,aps,prd,twocolumn,notitlepage,nobibnotes,nofootinbib,longbib,superscriptaddress]{revtex4-2}
\pdfoutput=1
\usepackage[utf8]{inputenc}
\usepackage[english]{babel}
\usepackage{standalone}
\usepackage{graphicx}
\usepackage{bm}
\usepackage{braket}
\usepackage{url}
\usepackage[colorlinks,citecolor=blue,linkcolor=blue,urlcolor=blue, breaklinks=true]{hyperref}
\usepackage{orcidlink}
\usepackage{bbold}
\graphicspath{{./figs}}
\usepackage{xcolor}
\usepackage{xspace}
\usepackage{ifthen}
\newcommand{\showcomments}{true} 

\newcommand{\ofek}[1]%
{\ifthenelse{\equal{\showcomments}{true}}%
{{\color{purple}{\small \textbf{Ofek:} #1}}}{\xspace}}%

\newcommand{\m}[1]%
{\ifthenelse{\equal{\showcomments}{true}}%
{{\color{blue}{\small \textbf{m:} #1}}}{\xspace}}%

\newcommand{\caslav}[1]%
{\ifthenelse{\equal{\showcomments}{true}}%
{{\color{orange}{\small \textbf{Časlav:} #1}}}{\xspace}}%

\newcommand{\Hc}{\mathcal{H}}
\newcommand{\Mc}{\mathcal{M}}
\newcommand{\hphi}{\hat\varphi}

\begin{document}

\title{Quantum Permutations and Beyond Quantum Controlled Reference Frames}
\date{\today}
\author{Ofek Bengyat\,\orcidlink{0000-0001-5547-9176}}
\email{ofek.bengyat@oeaw.ac.at}
\affiliation{Institute for Quantum Optics and Quantum Information (IQOQI) Vienna, Austrian Academy of Sciences, Boltzmanngasse 3, A-1090 Vienna, Austria}
\affiliation{University of Vienna, Faculty of Physics, Vienna Center for Quantum Science and Technology (VCQ),   Boltzmanngasse 5, A-1090 Vienna, Austria}

\author{Časlav Brukner}
\affiliation{Institute for Quantum Optics and Quantum Information (IQOQI) Vienna, Austrian Academy of Sciences, Boltzmanngasse 3, A-1090 Vienna, Austria}
\affiliation{University of Vienna, Faculty of Physics, Vienna Center for Quantum Science and Technology (VCQ),   Boltzmanngasse 5, A-1090 Vienna, Austria}

\author{Marios Christodoulou\,\orcidlink{0000-0001-6818-2478}}
\affiliation{Institute for Quantum Optics and Quantum Information (IQOQI) Vienna, Austrian Academy of Sciences, Boltzmanngasse 3, A-1090 Vienna, Austria}

\begin{abstract}
Quantum permutations, or magic unitaries, have in recent years been explored in the context of identifying `genuinely quantum' isometries of graphs. Here, we import this tool in physics, showing that quantum permutations yield a generalisation of quantum reference frames in a discrete setting that is reminiscent of the passage from special to general relativity. We show that the typical quantum reference frames framework corresponds to quantum permutations classified as `classical' in the mathematical literature, and which we demonstrate are quantum controlled transformations (superpositions of classical coordinate maps). Genuinely quantum permutations (i) allow to construct \emph{non--commuting} quantum reference frames (ii) correspond to \emph{local}, as opposed to global,  superpositions of transformations. Strikingly, we find that the non-commutativity of quantum fields, when used as reference systems, is exactly what implies that the change of frame is achieved through a genuine quantum permutation.
We illustrate the above with several examples in both first and second quantization formalism, which demonstrate (a) simultaneous control on non--commuting variables, (b) the existence of bipartite states that can be localized with a genuine quantum permutation and cannot be localized with the usual quantum reference frame transformations (without introducing additional degrees of freedom),  (c) extension of the Ising model symmetries to genuinely quantum permutations, and (d) extension of the symmetries of a scalar field action on curved spacetime to genuinely quantum permutations. While we have in mind applications in quantum gravity, we expect our formalism to be of interest in a wide range of topics in quantum information.

\end{abstract}
\maketitle

\section{Introduction}

Coordinates have played a crucial role in the development of modern physics. 
A choice of coordinates is made in order to describe a physical system. We may interpret this choice as a relational description of a physical system of interest with respect to another physical system---which realizes the coordinate system--- a system of reference. If we demand that all systems obey quantum mechanics, we are led to a description of a quantum system with respect to another quantum system. 

This is the logic developed in the program of quantum reference frames, with preliminary ideas appearing since the 80s \cite{aharonovQuantumFramesReference1984,rovelliQuantumReferenceSystems1991,Rovelli:1995fv}. Recent developments have used quantum information tools, e.g.~\cite{Loveridge_2018,giacominiQRF2019,zychRelativityQuantumSuperpositions2018,Vanrietvelde2020changeof,Castro-Ruiz_2020,hohnTrinityRelationalQuantum2021,hohnEquivalenceApproachesRelational2021,delahamettePerspectiveneutralApproachQuantum2021a,AliAhmad_2022}. Among the questions motivating this program, Ref.~\cite{giacominiQRF2019} asked how the world appears from the perspective of a laboratory that is in a quantum superposition relative to another laboratory, and whether the laws of physics are covariant under transformations between quantum reference frames. These developments were also informed by the parallel studies of quantum correlations with indefinite causal order e.g.~\cite{Chiribella:2009lvz,Oreshkov:2011er,PhysRevX.8.011047}.

While the applicability of a framework of quantum reference frames is potentially very wide, its development is especially oriented toward applications in quantum gravity. The system which serves as the reference, can be taken to be spacetime itself. In a quantum theory of gravity, spacetime itself must obey the laws of quantum physics. Then, the theory should come with a relational understanding of how to describe quantum matter with respect to a quantum spacetime, as there isn't anything else `behind spacetime' to use as reference. Similarly to how Einstein posited that physical laws should be invariant under changes of coordinate systems, it is natural to expect that in a quantum theory of spacetime background independence will be implemented as the invariance under changes of some kind of quantum coordinates. 

 Coordinates can be tied to the idea of a reference system, for instance, by taking the map that defines the coordinates to correspond to something of physical substance that takes different values at different spacetime locations---a test field. Considering this to be a quantum field, leads to a notion of quantum coordinates \cite{kabelQuantumCoordinatesLocalisation2025a}, an idea explored also in the context of attempts at formulating a quantum equivalence principle \cite{Zych:2015fka,Hardy:2019cef,giacominiQuantumSuperpositionSpacetimes2022}.

However, the notions of quantum reference frames and quantum coordinates which have appeared in the literature seem to correspond to a small subset of the possibilities: they correspond to quantum superpositions of \emph{classical} transformations, formed by quantum controlling on a reference system; in this sense, they are not `fully' quantum. How to go beyond, while maintaining that the transformations are unitary, has not been obvious. 

In this work, we show that the mathematical theory of quantum permutations is remarkable in that (i) it shows how to achieve concretely unitary reference frame transformations that are more general than quantum superpositions of classical transformations, and (i) this shortcoming has been noticed completely independently of physics. Tellingly, we will show that quantum permutations that are not of the quantum-controlled form have been termed `genuinely quantum' by mathematicians, reflecting the general intuition that truly quantum transformations should inherently involve non-commutativity---here, of the orthogonal projections that enter the definition of quantum permutations.

Quantum permutations, or magic unitaries, are studied in operator theory as a natural generalization of permutations to the quantum realm, see \cite{weberQuantumPermutationMatrices2023} for a review. The use of permutations implies that we will be working in a discrete setting. Here, we will focus on finite dimensional Hilbert spaces (although many of our results automatically extend to the countable case). Quantum permutations extend the classical symmetries of graphs to quantum symmetries: some pairs of graphs are classically not isomorphic, but are mapped to each other by a quantum permutation.  In this sense, these new symmetries are quantum symmetries, as they have no classical analogue. In the context of non-local games on graphs, it has been shown that  sometimes a classical winning strategy does not exist but only a quantum one, precisely when the graphs are not classically isomorphic but are quantum isomorphic  \cite{atseriasQuantumNonsignallingGraph2019}. Here, we import these striking results in physics using the language of quantum information. 

We will see that \emph{quantum permutations arise naturally in quantum theory as transformations between relational descriptions, and allow for local quantum reference frame choices that are point-wise non-commuting}: the orthogonal projections through which the transformation is defined at one point need not commute with those through which it is defined at another point. Interestingly, quantum coordinate transformations of this kind contain the usual quantum-controlled transformations (that are superpositions of classical transformations) as a proper subset, and it is the more general class that we will discuss in detail.

\bigskip

The material is organized as follows. We show that quantum permutations
are the transformations that allow to switch between descriptions of quantum systems with respect to other quantum systems. In standard quantum mechanics the quantum permutations act on the elements of the orthonormal basis defined by some relational observable (Section \ref{sec:1stq}). In the case of discrete quantum fields, quantum permutations act on locations (Section \ref{sec:qft}). Crucially, the non-commutativity of a reference quantum field with itself at two different points implies that the change of frame is achieved through a quantum permutation.

The above motivate importing the technology of quantum permutations to physics. For this, we translate them in the language of quantum information, where they are naturally interpreted as acting on a state space composed of two factors, the systems of interest and the reference systems (Section \ref{sec:math}). We show that the quantum permutations known as `classical' in the mathematical literature correspond to quantum controlled transformations (Section \ref{sec:qc}). This demonstrates in particular that the quantum permutations known as `genuinely quantum' in the mathematical literature are those that cannot be cast in a quantum controlled form. Hereafter, we introduce the physics oriented terminology quantum controlled (QC) permutations for what are known as the `classical' quantum permutations and beyond quantum controlled (BQC) permutations for those called `genuine' quantum permutations. We explore within our formalism the properties of building blocks of BQC permutations in Section \ref{sec:bqc}. 

Strikingly, the introduction of the point-wise non-commutativity of the transformations can be understood as allowing for local instead of only global quantum coordinate changes: BQC permutations correspond to \textit{local} superpositions of transformations, while QC permutations are \textit{global} superpositions of transformations (Section \ref{sec:localRelationality}). Any BQC permutation can be understood as making an independent choice of basis for quantum control at each point; taken together, these choices yield a transformation that is not QC, since arbitrary local basis choices will generically fail to commute.  \emph{This suggests that the passage from QC to BQC transformations allows for a generalisation of quantum reference frames akin to the passage from global coordinate changes to local coordinate changes, reminiscent of the passage from special to general relativity. }

To build intuition on BQC permutations, we examine an example of a genuinely quantum graph symmetry using a simple physical model of a spin on a graph (Section \ref{sec:toygraph}). This motivates understanding quantum permutations as symmetries of the dynamics, even when a graph structure is not present. We show that usual quantum reference frame transformations studied in the literature, which are symmetries of Hamiltonians of certain kinds, can be cast as QC permutations (Section \ref{sec:qrfs}). We extend the quantum reference frames framework using BQC permutations by giving examples of (a) transformations and Hamiltonians that are not quantum controlled (Section \ref{sec:bqchamiltonian}), (b) a bipartite state that can be localized with a BQC permutation and cannot be localized with a QC permutation (Section \ref{sec:localizetwo}), and (c) compare with related relevant literature in Section \ref{sec:literature}.
In Section \ref{sec:ising}, we see that the above imply the existence of new, genuinely quantum, symmetries of the Ising model. Finally, in Section \ref{sec:scalar} we show that quantum permutations that are differentiable in a discrete sense, are new, genuinely quantum, symmetries of the discretised action of a scalar field on a curved spacetime.

\section{Quantum permutations are necessary for relational descriptions among quantum systems}\label{sec:first}

We begin by a demonstration that quantum permutations (introduced formally in the following Section) emerge naturally as reference system transformations when considering quantum theory in a \emph{discrete and relational} setting. We first discuss (A) first-quantization systems and then (B) quantum fields. We will see that the role of QPs as reference system transformations or a sort of quantum coordinate maps becomes especially clear when considering quantum fields. 

\subsection{Quantum permutations from quantum mechanics}\label{sec:1stq}

Consider the standard first-quantization picture of a system $S$ described by a Hilbert space $\Hc_S$ and an observable (Hermitian operator) $\hat X_1$ on $\Hc_S$. In a relational approach, the eigenvalues of the observable $\hat X_1$ are not absolute, but, given in relation to some reference system $O_1$. In turn, assuming the reference to be a quantum system as well, it comes with its own Hilbert space $\Hc_{O_1}$. The standard first-quantization picture can then be understood as the system and reference described by a joint state space  $\Hc_S \otimes \Hc_{O_1}$, with the observable $\hat X_1$ extended trivially (as the identity) on the sector $\Hc_{O_1}$. That is, $\hat X_1$ is understood as $\hat X_1 \otimes \mathbb{1}$ acting on $\Hc_S \otimes \Hc_{O_1}$.

Now, we ask: \emph{what is the description of the physical property of $S$ corresponding to $\hat X_1$, if the reference system is not $O_1$ but some other quantum system $O_2$ with a state space $\Hc_{O_2}$}? This will correspond to some Hermitian $\hat X_2$,
acting on $\Hc_S \otimes \Hc_{O_2}$, and will not act trivially on $\Hc_{O_2}$. This is because it will depend on \emph{both} the (eigen)values of $\hat X_1$ \emph{and} the state of $O_2$.\footnote{The two descriptions coexist: as made explicit below, both $\hat X_1$ and $\hat X_2$ are observables on the same total space $\Hc_S \otimes \Hc_{O_1} \otimes \Hc_{O_2}$, each extended trivially on the sector it does not act upon. What distinguishes the two perspectives is therefore not the factorization of the Hilbert space, but which observables each reference has access to: $O_1$ has access to $\hat X_1$ and $O_2$ to $\hat X_2$. In the context of the discussions related to the Wigner's friend paradox, this implies that we are taking here the `God's eye view'. }

We now formally define the above and show that the general form of $\hat X_2$ is a conjugation of $\hat X_1$ by a quantum permutation. To avoid confusion, we stress that $\hat X_1$ and $\hat X_2$ refer to the \emph{same} relational physical quantity, for instance a relative position, as seen from the perspective of two different reference frames or observers, $O_1$ and $O_2$.

Let $S$ be a system of interest with \(\Hc_S\) its \(N\)-dimensional Hilbert space, and $\hat X_1$ an observable on \(\Hc_S\) with eigenvalues $x_1=x_1^{(1)},...,x_1^{(N)}$, given with respect to a reference system $\Hc_{O_1}$. Let $O_2$ be an additional reference system and \(\Hc_{O_2}\) Hilbert space. The total Hilbert space is thus $\Hc_S \otimes\Hc_{O_1} \otimes\Hc_{O_2}$. \textit{A relative observable with respect to the reference $O_2$}, describing the same physical quantity as $\hat X_1$, is an observable $\hat X_2$ on $\Hc_S \otimes \Hc_{O_2}$ with eigenvalues $x_2=x_2^{(1)},...,x_2^{(N)}$ such that (1) For every $x_1$ there exists an observable $\hat X_{2,x_1}$ such that $\hat X_2\ket {x_1} \ket\phi = \ket {x_1} \otimes \hat X_{2,x_1} \ket\phi$ for all $\ket\phi\in\Hc_{O_2}$, and (2) 
    For all $x_1\neq x_1'$, $\ket\phi ,\ket{\phi'} \in\Hc_{O_2}$ and $x_2$, if $\hat X_2 \ket {x_1} \ket\phi = x_2 \ket {x_1} \ket\phi$ and $\hat X_2 \ket {x_1'} \ket{\phi'} = x_2 \ket {x_1'} \ket{\phi'}$, then $\braket{\phi\vert\phi'}=0$.

These two simple conditions ensure that $X_2$ is a `good' relational observable. Condition 1 ensures that on eigenstates of $\hat{X}_1$, the corresponding (relative to $O_2$) observable $\hat X_2$ only acts non-trivially on the state of $O_2$. Intuitively, that when the state of $S$ from the point of view of $O_1$ is definite, whether it becomes indefinite after the transformation depends only on the state of $O_2$.

 Condition 2 ensures that eigenvalues relative to $O_2$ distinguish eigenvalues relative to $O_1$. In particular, it states that if two different eigenvalues of $\hat{X}_1$  are transformed to the same eigenvalue of $\hat{X}_2$ , this is only due to the states of $O_2$ being distinguishable. Intuitively, this condition ensures that no information about the system is lost when changing frames (it is analogous to the bijectivity assumption of the following sub-Section).

Now, it can be shown that any observable (Hermitian) $\hat X_2$ satisfies conditions 1 and 2 (it is a relative observable with respect to $O_2$) if and only if
\begin{align} \label{eq:x2=ux1}
    \hat X_2= u \; X_1 \; u^\dagger
\end{align}
for some
\begin{align}\label{eq:physics_form}
    u = \sum_{x_1,x_2} \ket {x_1} \bra {x_2} _S \otimes  u_{x_1x_2,\;O_2}.
\end{align}
satisfying\footnote{Note a subtlety: $\hat X_2$ is invariant to compositions of $u$ with arbitrary unitaries on the reference sector $\Hc_{O_1} \otimes \Hc_{O_2}$ (which act trivially on $S$). Nevertheless, the essence of the transformation is captured in the quantum permutation $u$, which acts non-trivially on the system of interest $S$.}
\begin{align}
\label{def:qp}
u_{x_1x_2} = &u_{x_1x_2}^\dagger = u_{x_1x_2}^2\\\nonumber 
\sum_{x_2} u_{x_1x_2}&=\sum_{x_1} u_{x_1x_2} = \mathbb1_{O_2}.
\end{align}
The proof is given in Appendix \ref{ap:proof1stq}. Condition 1 gives the first set of equations and the decomposition of identity with respect to $x_2$. Condition 2 yields the decomposition of identity with respect to $x_1$. As we discuss in the next Section, \eqref{def:qp} is the definition of a \emph{quantum permutation} $u$.

To give an example, consider \(S\) to be a particle on a line with \(\hat X_1\) its position relative to some reference observer (e.g.~an experimenter or apparatus) $O_1$, and $O_2$ another observer on the same line. The relative observable $\hat X_2$ distinguishes the locations of $S$ as measured by $O_2$. Concretely, take $\hat X_2$ to be the signed distance between $S$ and $O_2$, that is, $\hat X_2 = \hat X_1 - \hat X_{O_2}$, where $\hat X_{O_2}$ is the position of $O_2$ relative to $O_1$. Given two states with the same $\hat X_2$ eigenvalue but different $\hat X_1$ eigenvalues we have $\braket{\phi\vert\phi'}=0$ because $\ket\phi$ and $\ket{\phi'}$ are eigenstates of $\hat X_{O_2}$ with different eigenvalues (satisfying Condition 2). Anticipating what follows, note this simple example concerns a special  case of quantum permutations, a quantum controlled (QC) transformation. It is extended to an example of a beyond quantum controlled (BQC) transformation in Section \ref{sec:bqchamiltonian}.

We have seen that the defining properties of the quantum mechanical transformation between the relative to $O_1$ observable $\hat X_1$ and the relative to $O_2$ observable  $\hat X_2$, is a quantum permutation. This follows from simple conditions that assure a well defined relational description between quantum systems.

\subsection{Quantum permutations from quantum fields}\label{sec:qft}

We now consider that both the system of interest and the reference systems with respect to which it is described, are an abstract sort of discrete `quantum fields'. Here, by quantum field we simply mean a family of observables $\hphi(q)$ where $q$ are elements of some abstract finite set $\mathcal{M}$. No further structure is assumed in the formal setting of this Section. In later Sections, we take $q$ to be the nodes of a graph or lattice, $q$ play the role of a discrete analogue of the points of a spacetime manifold. The reader is forewarned that below whenever we write $\hat{\varphi}_1$ and  $\hat{\varphi}_2$, we do not mean that there are two fields, but that this is the same abstract field $\hat{\varphi}$ expressed in different (quantum) coordinates. Similarly, when we write $x_1$ and $x_2,$ we refer to different coordinates of some abstract point $q$. Also, note that in the previous sub-Section we did not have an analogue of the abstract field $\hat{\varphi}$ (which is defined on abstract points $q$), as in quantum mechanics there is no analogue of the abstract set $\mathcal{M}$.

Consider three quantum fields $\hphi(q)$, $\hat\chi_{1}(q)$ and $\hat\chi_{2}(q)$. These describe, respectively, the system of interest $\hphi$ and two reference systems $O_1,O_2$. All three fields are taken to be physical, two of them serving as references. We define the \emph{relative field} $\hphi_i(x^\mu)$ as describing $\hphi$ relative to $\hat\chi_i$, where $x^\mu$ is an eigenvalue of $\hat\chi_i$. Namely, the operator $\hphi_i(x^\mu)$ will coincide with the operator $\hphi(q)$ on states where the reference field $\hat\chi_i(q)$ takes the value $x^\mu$. 

More precisely, consider a quantum field $\hphi(q)$ that takes arbitrary states in some (Fock) space $\Hc_\varphi$. This is the system of interest. Consider also two reference quantum fields $\Hc_{O_i}$ for $i=1,2$, each composed of 4 quantum scalar fields $\hat\chi_i(q) = \big( \hat\chi_i^{(\mu)}(q) \big)_{\mu=0,1,2,3}$ for $q\in\Mc$. The spectra of the operators $\hat\chi_i(q)$ are thus points $x^\mu \in \mathbb R^4$. The total Hilbert space is $\Hc = \Hc_\varphi \otimes \Hc_{O_1} \otimes \Hc_{O_2} $. By a  relative to $O_i$ quantum field $\hphi_i$, we mean that there is a (discrete) subset of $\mathbb R^4$ such that for all $\ket\psi\in\Hc$ with $\hat\chi_i(q)\ket\psi = x^\mu \ket\psi$, it holds that $\hphi_i(x^\mu)\ket\psi = \hphi(q)\ket\psi$. 

For $\hphi_i$ to be well defined, $O_i$ need to be good reference systems. For this, we restrict to subsets of $\Mc$ and $\Hc_{O_i}$ such that $\Hc_{O_i}$ only contain states of the reference fields that have enough variability to uniquely name all points in $\Mc$. This means states for which the reference field is `bijective' in the following sense: there exists a family of operators $\hat \chi_i^{-1}(x^\mu)$ on the spectrum of all $\hat\chi_i(q)$ such that $\hat\chi_i(q)\ket{\phi_i} = x^\mu \ket{\phi_i}$ if and only if $\hat \chi_i^{-1}(x^\mu) \ket{\phi_i} = q \ket{\phi_i}$.\footnote{Since $q$ is generally not a complex number, we abuse the notation $q\ket{\phi_i}$ to mean $\lambda_q\ket{\phi_i}$, with $\lambda_q$ some fixed injection of $\Mc$ into $\mathbb C$.} In a sense, this is a simple `quantum version' of coordinate maps. Note that relative fields $\hphi_i(x^\mu)$ would be analogous to the usual fields of quantum field theory if  $\hat\chi(q)$ were classical coordinate maps.

Now, define $P_i^{qx}$ to be the orthogonal projection on the $x^\mu$ eigenspace of $\hat\chi_i(q)$. It holds that $\sum_{x^\mu} P_i^{qx} = \mathbb1_{O_i}$,  $P_i^{qx} = P_i^{qx,\dagger}$ and $P_i^{qx} P_i^{qx'} = \delta_{xx'} P_i^{qx}$. Crucially, due to the `bijectivity' assumption that $O_i$ are good reference quantum fields, $P_i^{qx}$ is then also the orthogonal projection on the $q$ eigenspace of $\hat \chi_i^{-1}(x^\mu)$. This means that it \emph{also} holds that $\sum_q P_i^{qx} = \mathbb1_{O_i}$. That is, a summation on any of the upper indices of the orthogonal projections $P_i^{qx}$ yields the identity. This is the characteristic property of quantum permutations.

Let us find the field $\hphi_i(x^\mu)$ explicitly. For all $q\in\Mc$ and $x^\mu$ we have,
\begin{align}\label{eq:phi_i}
    \hphi_i(x^\mu) P_i^{qx} = \hphi(q) \otimes P_i^{qx}.
\end{align}
Summing over $q$, we get $\hphi_i(x^\mu) \sum_q P_i^{qx} = \sum_q \hphi(q) \otimes P_i^{qx}$. From the bijectivity assumption it follows that $\sum_q P_i^{qx} = \mathbb{1}_{O_i}$, which yields the relative field
\begin{align}
\label{eq:relField}
    \hphi_i(x^\mu) = \sum_{q} \hphi(q) \otimes P_i^{qx} .
\end{align}
Then, we have
\begin{align}\label{eq:phi12rule}
    \hphi_2(x_2^\mu) = \sum_{x_1^\mu} \hphi_1(x_1^\mu)\; u_{x_2x_1}
    = \sum_{x_1^\mu} u_{x_2x_1}\;\hphi_1(x_1^\mu)
\end{align}
with
\begin{align}\label{eq:qpfield}
    u_{x_2x_1} = \mathbb1_S\otimes\sum_{q} P_1^{qx_1}  \otimes  P_2^{qx_2}.
\end{align}

As in Section \ref{sec:1stq}, the $u_{x_2x_1}$ satisfy
\begin{align}
u_{x_2x_1} &= u_{x_2x_1}^\dagger = u_{x_2x_1}^2\\\nonumber 
\sum_{x_1} u_{x_2x_1}&=\sum_{x_2} u_{x_2x_1} = \mathbb1_{O_1O_2},
\end{align}
which are the defining properties of a quantum permutation (see next Section).  
Therefore, with minimal assumptions on the reference systems to enable well-defined relative descriptions of otherwise arbitrary systems of interest, it follows that the transformation between relative fields is given by a quantum permutation.

To make contact with the formalism of the previous subsection where we discussed first-quantisation, note that another formal way to write the transformation is as the matrix multiplication:
\begin{align}
\begin{pmatrix}
\hphi_2(x_2^{(1)})\\
\hphi_2(x_2^{(2)})\\
\vdots\\
\hphi_2(x_2^{(N)})
\end{pmatrix}
=
\begin{pmatrix}
u_{x_2^{(1)}x_1^{(1)}} & \cdots & u_{x_2^{(1)}x_1^{(N)}}\\
u_{x_2^{(2)}x_1^{(1)}}  & \cdots & u_{x_2^{(2)}x_1^{(N)}}\\
\vdots  & \ddots & \vdots\\
u_{x_2^{(N)}x_1^{(1)}}  & \cdots & u_{x_2^{(N)}x_1^{(N)}}
\end{pmatrix}
\begin{pmatrix}
\hphi_1(x_1^{(1)})\\
\hphi_1(x_1^{(2)})\\
\vdots\\
\hphi_1(x_1^{(N)})
\end{pmatrix},
\end{align}
where recall that $x_i^{(k)}$ are the eigenvalues of the reference field $\hat\chi_i(q)$.

Now, let us make two important remarks.
First, it is not only the transformation $u$ that fulfills the definition of a quantum permutation (QP), but also the sets of orthogonal projections $P_i$. While the transformation $u$ transforms between the fields relative to each reference  --- $\hphi_1$ and $\hphi_2$ --- the operators $P_i^{qx}$ encode the relation between the abstract field $\hphi$ and each of the relative fields $\hphi_i$. In this sense, the $P_i$ can be thought of as a quantum analogue to coordinate maps, and the quantum permutation $u$ as a quantum coordinate change (it encodes how to switch between coordinates assigned by $O_1$ and those assigned by $O_2$).\footnote{Interestingly, the relation \eqref{eq:qpfield} that produces a QP $u$ from two QPs $P_1, P_2$ is a known operation in the mathematical literature called the Woronowicz product \cite{woronowiczCompactMatrixPseudogroups1987,weberQuantumPermutationMatrices2023}.}

  Second, if the reference fields do not commute at two points $q$ and $q'$, the corresponding orthogonal projectors will also not commute. That is, 
\begin{equation}
\label{eq:microcausality}
    [ \hat\chi_i(q), \hat\chi_i(q') ] =0 \Leftrightarrow [P_i^{qx_i}, P_i^{q'x_i'}] =0 \ \ \forall x_i,x'_i
\end{equation}
Since it is a characteristic property of quantum fields that they may not commute at different points (also known as the principle of micro-causality), this implies that generic quantum reference frame transformations will involve such non--commuting projectors. This implies that the quantum permutation that does changes from the frame of one field to the other must be `genuinely quantum'.

\section{Quantum permutations}\label{sec:math}

\begin{figure}
    \centering
    \includegraphics[width=0.8\linewidth]{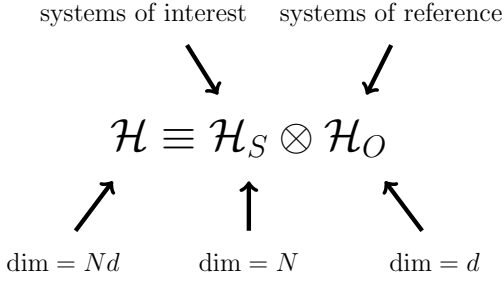}
    \caption{The names and dimensions of the Hilbert space on which magic unitaries act. We regard $\Hc_X$  as a system of interest. It contains the eigenbasis $\ket{x}$ of the observable $X$. This basis is the one permuted. $\Hc_O$ can be regarded as an external observer, as is the case in QRFs.}
    \label{fig:totalspace}
\end{figure}

Having seen that quantum permutations arise naturally when considering relational descriptions among quantum systems, we now give their precise definition (see \cite{weberQuantumPermutationMatrices2023} for further detail) and express them in the bra-ket notation. 

Take $\Hc_O$ a separable Hilbert space, whose dimension we denote $d$, which could be either finite or infinite. A \emph{quantum permutation}, or \emph{magic unitary} $u$ is a collection $(u_{xy})_{1\leq x,y \leq N}$ of bounded operators $u_{xy}\in \mathcal{B}(\Hc_O)$, with $N$ finite, which are orthogonal projections that partition $\Hc_O$ into orthogonal subspaces for a constant $x$ or $y$, arranged as follows. The operators $u_{xy}$ compose the quantum permutation $u$ by understanding the indexes $x,y$ as signifying rows and columns arranged in a `matrix of matrices' $u$ with entries $u_{xy}$. Note that while $u$ is a unitary, the `entries' $u_{xy}$ are orthogonal projections and generally not unitaries.

That the operators $u_{xy}$ are orthogonal projections is equivalent to the condition
\begin{align}
\label{eq:orthopro}
    u_{xy}^\dagger = u_{xy} = u_{xy}^2.
\end{align}
That the $u_{xy}$ partition the entire $\Hc_O$ in orthogonal subspaces is equivalent to demanding 
\begin{align}
\label{eq:identpart}
    \sum_z u_{xz} = \sum_z u_{zy} = \mathbb{1}_O \text{   } \ \ \forall x,y.
\end{align}
Then, since each $u_{xy}$ is to be understood as an entry in a matrix of matrices $u$, the above condition dictates that each row and each column sums up to the identity operator of $\Hc_O$.

It follows from the above definition that for $a\neq b$ and for all $x,y$, the pairs $u_{xa}$ and $u_{xb}$, as well as the pairs $u_{ay}$ and $u_{by}$,  are disjoint projections. That is, 
\begin{align}
\label{eq:disjoint}
    u_{xa}u_{xb} & = 0 \nonumber \\ u_{ay}u_{by} & = 0.
\end{align}
This is due to a more general property of projections that we will be using often: projections $p_i^2=p_i$ that sum up to the identity, $\sum_i p_i=\mathbb{1}$ are necessarily mutually orthogonal, $p_i p_j = 0$ for $i\neq j$. In particular, this implies that all entries of the same row or the same column of a quantum permutation commute. An example of a quantum permutation with four rows and columns is
\begin{align}\label{eq:eg1}
u = 
\begin{pmatrix}
\pi_1 & \mathbb{0}_O & \mathbb{1}_O - \pi_1 & \mathbb{0}_O \\
\mathbb{1}_O - \pi_1 & \mathbb{0}_O & \pi_1 & \mathbb{0}_O \\
\mathbb{0}_O & \pi_2 & \mathbb{0}_O & \mathbb{1}_O - \pi_2 \\
\mathbb{0}_O & \mathbb{1}_O - \pi_2 & \mathbb{0}_O & \pi_2
\end{pmatrix}
\end{align}
with $\pi_i$ orthogonal projections on $\Hc_O$.

Quantum permutations are a generalization of a permutation matrix in the following sense. A permutation matrix has all zeros and a single unit in any row and column. So, all rows and columns of a permutation matrix sum up to one. A quantum permutation retains and generalizes this feature: all rows and columns sum to the unit \emph{matrix}. For this reason, quantum permutations are also known as `magic unitaries' (since ordinary permutation matrices are `magic squares'). When the entries $u_{xy}$ are one by one matrices, the definition of a quantum permutation coincides with ordinary permutations. When all entries $u_{xy}$ commute among them, the quantum permutation is called `classical' and when at least one pair $u_{xy}$ and $u_{x'y'}$ does not commute it is called `genuine', for reasons that will become clear in the following Sections.

To understand the action of quantum permutations on quantum mechanical states, it is useful to express them in the bra-ket notation. By inspection of the definition of quantum permutations, they are operators on a Hilbert space of the form
\begin{align}
\label{eq:totalSpace}
   \Hc \equiv \Hc_S \otimes \Hc_O .
\end{align}
where $\Hc_S $ is $N$--dimensional, $\Hc_O $ is $d$--dimensional, and $\Hc$ is therefore of dimension $Nd$. Later on, the factor $\Hc_S $ will be the space of the systems of interest and the factor $\Hc_O$ the space of the reference systems or `observers', see Figure \ref{fig:totalspace}. In the next Section, where we consider the special case of quantum controlled transformations, $\Hc_S $ is the target space
and $\Hc_O$ the control space. 

To give an example of the mechanics of the formalism, consider any state $\ket\psi\in\Hc$. This can be written as
\begin{align}
\ket\psi =\sum_{x=1}^N  \psi(x) \ket{x}\ket{\phi_x} \in \Hc
\end{align}
with  $\ket{x}\in\Hc_S$ an eigenvector of some Hermitian $\hat{X} \otimes \mathbb{1}$, $\psi(x)$ a (discrete) normalised wavefunction, and $\ket{\phi_x}$ some arbitrary normalized vector in $\Hc_O$. In the terminology of Section \ref{sec:1stq}, $\hat{X}$ is a relative observable corresponding to the observer with respect to which the tensor product structure was defined (as it acts trivially on $\Hc_O$), and $\ket{\phi_x}$ is some joint state of the reference systems. Applying $u$ results in
\begin{align}\label{eq:uname}
    u\ket\psi =  \sum_x  \ket{x}  \otimes \sum_{x'} \psi(x') \; u_{xx'}\ket{\phi_x'}.
\end{align}
In matrix form, this reads
\begin{align}
=
    \begin{pmatrix}
        u_{11}&\dots&u_{1N}\\
        \vdots&     & \vdots\\
        u_{N1}&\dots& u_{NN}
    \end{pmatrix}
    \begin{pmatrix}
        \psi(1)\ket{\phi_1}\\
        \vdots\\
        \psi(N)\ket{\phi_N}
    \end{pmatrix}
   .
\end{align}
A physics-oriented notation for writing a quantum permutation that we found convenient in calculations is \eqref{eq:physics_form}, repeated here for the reader's convenience
\begin{align}
    u = \sum_{x,x'} \ket {x} \bra {x'} \otimes  u_{xx'}\nonumber.
\end{align}
Having reviewed their definition and basic properties, in the following two Sections we investigate how quantum permutations act on quantum mechanical states.

\section{QUANTUM CONTROLLED PERMUTATIONS}\label{sec:qc}

The simplest extension of the concept of permutations to the quantum realm is to consider quantum controlled permutations. These are superpositions of classical permutations. Assume $\sigma_k$ are $K$ classical permutations matrices on $N$ items.\footnote{Note that $K$ is a different unrelated number than $N$ and $d$. The dimension of the total space is still taken to be $Nd$.} That is, $\sigma_k$ is an $N\times N$ matrix with a single unit entry in each row and column and remaining entries are zero. Given $\pi_k$ orthogonal projections $\pi_k^\dagger = \pi_k =\pi_k^2$ on $\Hc_O$ which sum up to the identity $\sum_k \pi_k = \mathbb{1}_O$, we define a \emph{quantum controlled (QC) permutation} as a unitary of the form
\begin{align}
\label{eq:qcontrolled}
    u = \sum_{k=1}^K   \sigma_k \otimes \pi_k.
\end{align}

The $\sigma_k \otimes \pi_k$ are matrices of matrices that look like the permutation matrix $\sigma_k$, but with the operator $\pi_k$ inserted in the place of the ones and the zero operator inserted in the place of the zeros. When restricted to any subspace $\pi_k \Hc_O$ of $\Hc_O$, the permutation is definite --- it is $\sigma_k$.  In this context, we can think of $\Hc_S$ from Section \ref{sec:1stq} as the target space and $\Hc_O=\Hc_{O_1} \otimes \Hc_{O_2}$ as the control space.

The subset of quantum controlled permutations can be characterized in a different way. It is immediate that all entries in a quantum permutation of the form \eqref{eq:qcontrolled} commute, as all $\pi_k$ commute. The opposite direction is not so straightforward but can be shown, we give the proof in Appendix~\ref{ap:proof}. That is, \emph{QC permutations are exactly all quantum permutations whose entries all commute.} This proves that QC permutations are the subset of QPs known as `classical QPs' in the mathematical literature.

Let us see this with a simple example. Consider a quantum permutation $u$ where the entries $u_{xy}$ are two by two matrices. In contrast to ordinary permutations that only have one non zero entry per row and column, it is now possible to have \emph{two} non zero entries: each will project on a one dimensional orthogonal subspace, and the sum of both is the two by two identity matrix. It is this behavior that allows one to encode the intuitive requirement, for example, of `permuting 1 with 2 in quantum superposition of permuting 1 with 3'. This superposition of classical permutations is the quantum permutation
\begin{align}
    u &= 
    \begin{pmatrix}
\mathbb{0}_O & \ket{\uparrow}\bra{\uparrow} & \ket{\downarrow}\bra{\downarrow} \\
\ket{\uparrow}\bra{\uparrow} &\ket{\downarrow}\bra{\downarrow} & \mathbb{0}_O \\
\ket{\downarrow}\bra{\downarrow} & \mathbb{0}_O & \ket{\uparrow}\bra{\uparrow}
\end{pmatrix}
\\\nonumber
&=
    \begin{pmatrix}
        0 & 1 & 0 \\
        1 & 0 & 0 \\
        0 & 0 & 1
    \end{pmatrix} \otimes \ket{\uparrow}\bra{\uparrow} 
    +
    \begin{pmatrix}
        0 & 0 & 1 \\
        0 & 1 & 0 \\
        1 & 0 & 0
    \end{pmatrix} \otimes \ket{\downarrow}\bra{\downarrow}
,
\end{align}
where $\Hc_O=\mathbb{C}^2$ is spanned by the orthonormal basis $\ket\uparrow,\ket\downarrow$.

The reason quantum permutations whose entries $u_{xy}$ commute is called a `classical' permutation in the context of graph symmetries \cite{weberQuantumPermutationMatrices2023} can be seen from \eqref{eq:qcontrolled}: every graph that has such a quantum controlled permutation as a symmetry will also have each of the ordinary permutations $\sigma_k$ as symmetries as well (it will have more than one classical symmetries, and the QC symmetry simply corresponds to a superposition of those). In the mathematical literature, it is only when any of the entries of the quantum permutation matrix \emph{do not commute} that we say that we have a `\emph{genuinely quantum permutation}'. For example, in \eqref{eq:eg1} it corresponds to the requirement that $[\pi_1,\pi_2] \neq 0$. For an explicit example of a  genuinely quantum symmetry of a graph see Section \ref{sec:toygraph}.

We have been presented with a nomenclature clash in mathematics and physics. Transformations of the quantum controlled form \eqref{eq:qcontrolled} are generally thought to be `quantum' in physics. We have shown that the quantum permutations defined by entries (orthogonal projectors) which are all commuting are exactly those of the form \eqref{eq:qcontrolled}. These have been called `classical permutations' in the mathematical theory of quantum permutations, because the `hallmark of quantumness' in that context is considered to be the non--commutativity of the orthogonal projectors composing the quantum permutation matrix. Instead of `classical permutations', we have termed them here quantum controlled (QC) permutations. Correspondingly, when at least two entries of the quantum permutation do not commute, instead of `genuinely quantum permutations' we introduce the terminology \emph{beyond quantum controlled (BQC) permutations}.

\section{BEYOND QUANTUM CONTROLLED PERMUTATIONS}\label{sec:bqc}

We examined so far the special case of quantum controlled permutations. These were a simple extension of classical transformations via the principle of quantum superposition. We now turn to understanding the general case of quantum permutations. 

Consider the following BQC permutation for $N=4$ and $d$ arbitrary\footnote{A technical counting argument yields that BQC permutations only exist for $N\geq 4$ \cite{weberQuantumPermutationMatrices2023}.} 
\begin{align}\label{eq:eg2}
u=
\begin{pmatrix}
\pi_1 & \mathbb{1}_O-\pi_1 & \mathbb{0}_O & \mathbb{0}_O \\
\mathbb{1}_O-\pi_1 & \pi_1 & \mathbb{0}_O & \mathbb{0}_O \\
\mathbb{0}_O & \mathbb{0}_O & \pi_2 & \mathbb{1}_O-\pi_2 \\
\mathbb{0}_O & \mathbb{0}_O & \mathbb{1}_O-\pi_2 & \pi_2 \\
\end{pmatrix},
\end{align}
with $\pi_1 \pi_2 \neq \pi_2 \pi_1 $. Recall that in order for this to be a quantum permutation it also holds that $\pi_k=\pi_k^2=\pi_k^\dagger$. 
BQC permutation of the form \eqref{eq:eg2} are building blocks of more general BQC permutations, much like any permutation group can be written in terms of 2-cycles \cite{Jung_2020}.

As discussed in the previous Section and proved in Appendix~\ref{ap:proof}, any quantum permutation with at least two non-commuting entries \emph{cannot} be put in the form~\eqref{eq:qcontrolled}. Therefore, this quantum permutation cannot be understood as a controlled application of classical permutations on the $N=4$ elements it acts upon. 
 
Now, we can rewrite $u$ as 
\begin{align}
\label{eq:eg2expanded}
    u &=
    \begin{pmatrix}
    \mathbb{1}_2 & \mathbb{0}_2  \\
    \mathbb{0}_2  & \mathbb{0}_2 \\ 
    \end{pmatrix}
    \otimes \pi_1 
    +
    \begin{pmatrix}
    \sigma & \mathbb{0}_2  \\
    \mathbb{0}_2  & \mathbb{0}_2 \\ 
    \end{pmatrix}
    \otimes (\mathbb{1}_O - \pi_1 )
    \\\nonumber
    &+
    \begin{pmatrix}
    \mathbb{0}_2  & \mathbb{0}_2 \\
    \mathbb{0}_2  & \mathbb{1}_2 \\ 
    \end{pmatrix}
    \otimes \pi_2
    +
    \begin{pmatrix}
     \mathbb{0}_2& \mathbb{0}_2  \\
    \mathbb{0}_2  & \sigma \\ 
    \end{pmatrix}
    \otimes (\mathbb{1}_O - \pi_2 )
    ,
\end{align}
where $\sigma = \begin{pmatrix}
        0&1\\
        1&0\\
    \end{pmatrix}$
is the $2\times 2$ non-trivial permutation. We see that $u$ is a sum of four terms; in each term, the left factor is acting on the $4$--dimensional $\Hc_S$ and the right factor is acting on the $d$--dimensional $\Hc_O$. Written in this way, it resembles a situation where we think of $\Hc_O$ as the control space. 

Recall that in the usual quantum information notion of a quantum controlled transformation, we would have a sum over terms where the left factors are unitaries on a target space $\Hc_S$. The right factors would be orthogonal projections that sum up to the identity on the control space $\Hc_O$. 

However, there are two important differences when considering a BQC permutation. First, in \eqref{eq:eg2expanded} the operators acting on the target space $\Hc_S$ are \emph{not} unitaries on $\Hc_S$ --- they are unitaries on \emph{subspaces} of $\Hc_S$. Second, the control space $\Hc_O$ is not acted upon with complementary projections. The pair $\pi_1$ and $\mathbb{1}_O-\pi_1$ sum to the identity, as does the pair $\pi_2$ and $\mathbb{1}_O-\pi_2$. However, \emph{all} four of them appear in \eqref{eq:eg2expanded}, while, for instance, $\pi_1$ and $\pi_2$ do \emph{not} commute. 

Intuitively, \eqref{eq:eg2expanded} can be thought of as \emph{two} quantum controlled transformations packaged together in one unitary, which is not overall a quantum controlled transformation. The first two terms correspond to a (unitary) quantum controlled operation on the subspace given by restricting to the `upper' half of $\Hc_S$ and the last two terms correspond to a (unitary) quantum controlled operation on the subspace given by restricting to the `lower' half of $\Hc_S$. 

It is this insight that yields the remarkable property that two unitary operators sum up to a unitary: in general, the sum of two unitaries is not a unitary. In the next Section, we see how this can be used to construct arbitrary BQC transformations. We will see in the next Section that in a certain sense this allows to construct local superpositions of transformations (while the quantum controlled case corresponds to global superpositions of transformations).

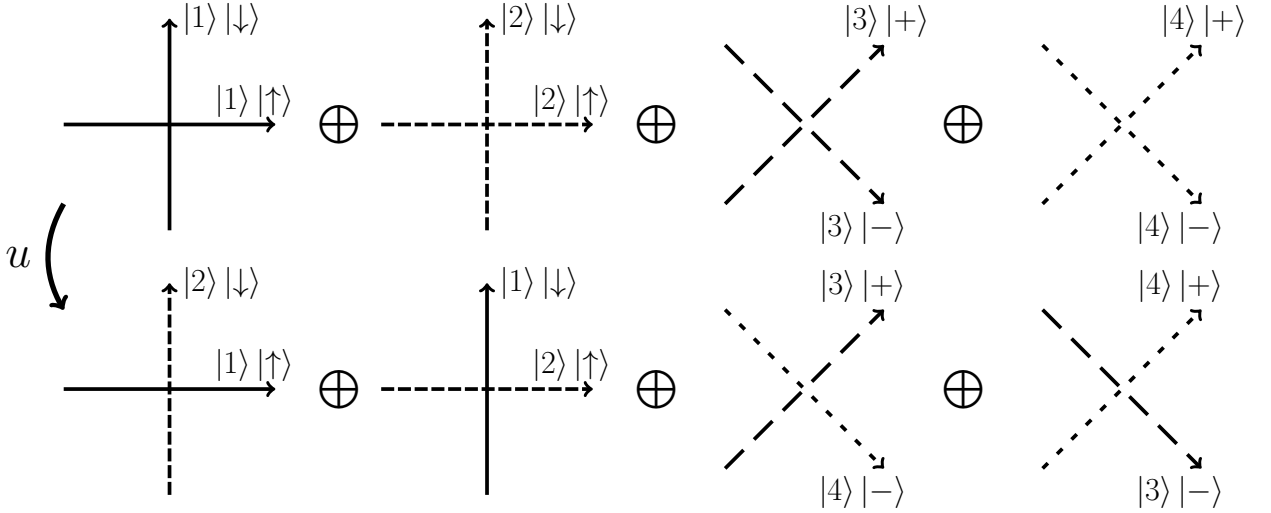
\begin{figure*}
    \centering
    \newcommand{\thickness}{2pt}
\tikzset{
    thickstyle/.style={->, line width=\thickness},
    dashedstyle/.style={->, line width=\thickness, dash pattern=on 14pt off 8pt},
    dottedstyle/.style={->, line width=\thickness, dash pattern=on 7pt off 2pt},
    thindotstyle/.style={->, line width=\thickness, dash pattern=on 4pt off 7pt},
}


\begin{tikzpicture}
    \draw[thickstyle] (-2,0) -- (2,0) node[above, font=\bfseries\LARGE] {\hspace{-0.8cm}$\ket1\ket\uparrow$};
    \draw[thickstyle] (0,-2) -- (0,2) node[right, font=\bfseries\LARGE] {\hspace{0.1cm}$\ket1\ket\downarrow$};

    \node at (3.2, 0) {\huge $\mathord{\bigoplus}$};

    \begin{scope}[xshift=6cm]
        \draw[dottedstyle] (-2,0) -- (2,0) node[above, font=\bfseries\LARGE] {\hspace{-0.8cm}$\ket2\ket\uparrow$};
        \draw[dottedstyle] (0,-2) -- (0,2) node[right, font=\bfseries\LARGE] {\hspace{0.1cm}$\ket2\ket\downarrow$};
    \end{scope}

    \node at (9.2, 0) {\huge $\mathord{\bigoplus}$};

    \begin{scope}[xshift=12cm]
        \draw[dashedstyle] (-1.5,1.5) -- (1.5,-1.5) node[below, font=\bfseries\LARGE] {\hspace{-0.8cm}$\ket3\ket{-}$};
        \draw[dashedstyle] (-1.5,-1.5) -- (1.5,1.5) node[above, font=\bfseries\LARGE] {\hspace{0.1cm}$\ket3\ket{+}$};
    \end{scope}

    \node at (15, 0) {\huge $\mathord{\bigoplus}$};

    \begin{scope}[xshift=18cm]
        \draw[thindotstyle] (-1.5,1.5) -- (1.5,-1.5) node[below, font=\bfseries\LARGE] {\hspace{-0.8cm}$\ket4\ket{-}$};
        \draw[thindotstyle] (-1.5,-1.5) -- (1.5,1.5) node[above, font=\bfseries\LARGE] {\hspace{0.1cm}$\ket4\ket{+}$};
    \end{scope}

    \draw[thickstyle] (-2,-5) -- (2,-5) node[above, font=\bfseries\LARGE] {\hspace{-0.8cm}$\ket1\ket\uparrow$};
    \draw[dottedstyle] (0,-7) -- (0,-3) node[right, font=\bfseries\LARGE] {\hspace{0.1cm}$\ket2\ket\downarrow$};

    \node at (3.2, -5) {\huge $\mathord{\bigoplus}$};

    \begin{scope}[xshift=6cm]
        \draw[dottedstyle] (-2,-5) -- (2,-5) node[above, font=\bfseries\LARGE] {\hspace{-0.8cm}$\ket2\ket\uparrow$};
        \draw[thickstyle] (0,-7) -- (0,-3) node[right, font=\bfseries\LARGE] {\hspace{0.1cm}$\ket1\ket\downarrow$};
    \end{scope}

    \node at (9.2, -5) {\huge $\mathord{\bigoplus}$};

    \begin{scope}[xshift=12cm]
        \draw[dashedstyle] (-1.5,-6.5) -- (1.5,-3.5) node[above, font=\bfseries\LARGE] {\hspace{-0.8cm}$\ket3\ket{+}$};
        \draw[thindotstyle] (-1.5,-3.5) -- (1.5,-6.5) node[below, font=\bfseries\LARGE] {\hspace{-0.8cm}$\ket4\ket{-}$};
    \end{scope}

    \node at (15, -5) {\huge $\mathord{\bigoplus}$};

    \begin{scope}[xshift=18cm]
        \draw[thindotstyle] (-1.5,-6.5) -- (1.5,-3.5) node[above, font=\bfseries\LARGE] {\hspace{-0.8cm}$\ket4\ket{+}$};
        \draw[dashedstyle] (-1.5,-3.5) -- (1.5,-6.5) node[below, font=\bfseries\LARGE] {\hspace{-0.8cm}$\ket3\ket{-}$};
    \end{scope}

    \draw[->, line width = 1mm, bend right=30] (-2, -1.5) to (-2, -3.5);

    \node at (-2.5, -2.5) [left, font=\bfseries\Huge] {$u$};

\end{tikzpicture}
    \caption{An illustration of the action of the BQC permutation $u$ from \eqref{eq:eg2} with $\pi_1 =\ket\uparrow\bra\uparrow, \pi_2=\ket{+}\bra{+}$ on the Hilbert space $\Hc = \Hc_S \otimes \Hc_O = \bigoplus_{x=1}^4  \ket{x}\otimes \Hc_O$, with $\Hc_O = \mathbb C^2$. That is, $N=4$, $d=2$ and the total Hilbert space is of dimension $Nd=8$. The figure illustrates how the different subspaces of each original copy of $\Hc_O$ move from one component to the other, creating a different decomposition of $\Hc$ into orthogonal subspaces.}
    \label{fig:decomp}
\end{figure*}

Let us now calculate explicitly how a BQC permutation acts on a basis of $\Hc$ using a simple toy-model. Take $N=4$ and $d=2$. Imagine a particle described by a wavefunction giving an amplitude to be in any of four positions, and $\Hc_O$ to be the 2-dimensional state space of some reference system spanned by $\ket\uparrow,\ket\downarrow$. Consider the quantum permutation \eqref{eq:eg2} with $\pi_1 =\ket\uparrow\bra\uparrow$ and $\pi_2 =\ket+\bra+ $, with $\ket\pm := \frac{1}{\sqrt2}\Big( \ket\uparrow\pm\ket\downarrow\Big)$. This completely defines the quantum permutation. To understand its action, it is sufficient to examine how it acts on the eight elements of a basis of $\Hc$: 
\begin{align}
\label{eq:explbqc}
    u\ket1\ket\uparrow &= \ket1\ket\uparrow \nonumber \\
    u\ket2\ket\uparrow &= \ket2\ket\uparrow \nonumber \\
    u\ket1\ket\downarrow &= \ket2\ket\downarrow \nonumber \\
    u\ket2\ket\downarrow &= \ket1\ket\downarrow \nonumber  \\ \ \nonumber \\
    u\ket3\ket+ &= \ket3\ket+ \nonumber \\
    u\ket4\ket+ &= \ket4\ket+ \nonumber \\
    u\ket3\ket- &= \ket4\ket-  \nonumber \\
    u\ket4\ket- &= \ket3\ket-.
\end{align}

The left hand side above is $u$ acting on a basis element of $\Hc$ and the right hand side shows the resulting state. We see that $u$ exchanges the labels $1$ and $2$ when $O$ is in $\ket\uparrow$, and does nothing to them when it is in $\ket\downarrow$. A similar thing happens for $3$ and $4$, but in the $\pm$ basis instead of the $\uparrow\downarrow$ basis. We see here explicitly that the transformation of the first four basis elements corresponds to a quantum controlled operation on the subspace given by restricting to the `upper' half of $\Hc_S$ spanned by $\ket{1}$ and $\ket{2}$, and the action on the last four basis elements corresponds to a quantum controlled operation on the subspace given by restricting to the `lower' half of $\Hc_S$ spanned by $\ket{3}$ and $\ket{4}$. This example of a BQC permutation is depicted in Fig.~\ref{fig:decomp}. 

With these tools in hand, we now turn to the construction of arbitrary quantum permutations.

\section{Local relativity of superposition}
\label{sec:localRelationality}
In the previous Section we remarked that the matrix structure and properties of BQC permutations enables to appropriately sum quantum controlled unitaries, packaging them into larger unitaries. We will now build on this insight to show that \emph{BQC (QC) transformations are formed by local (global) superpositions of transformations.}

First, note that the quantum controlled of \eqref{eq:eg2} is block diagonal. This can be generalised to a recipe to construct quantum permutations, by the observation that transformations of the following form are unitaries:
\begin{equation}\label{eq:bqcblocks}
    u = \sum_{b=1}^B \sum_{k=1}^{K_b} \sigma_k^{(b)} \otimes \pi_k^{(b)}.
\end{equation}
Here, $\pi_k^{(b)}$ are orthogonal projections and for each $b$, $\sum_{k=1}^{K_b}\pi_k^{(b)}=\mathbb{1}_O $. For each $b$, the matrices $\sigma_k^{(b)}$ each contain a permutation on a block specific to $b$ and zeroes elsewhere. Each two such $b$-blocks are disjoint and together form a partition of $1,...,N$. The quantum permutation \eqref{eq:eg2expanded} is an example with $B=2,  K_1=K_2=2$.\footnote{This way to construct `genuinely quantum permutations' corresponds to the \emph{disjoint isomorphism criterion} found in \cite{levandovskyyExistenceQuantumSymmetries2022}.} This property will be used in Section \ref{sec:bqchamiltonian} to construct beyond quantum controlled quantum reference frames.

Second, note that when $B=N$, each `block' is $d$-dimensional, the dimension of the observers or reference sector $\mathcal{H}_O$. In this case, we can write any QP in the following form
\begin{equation}\label{eq:local_basis}
    u = \sum_{x}^N \sum_{k=1}^{K_x} \ket{y_k(x)}\bra{x} \otimes \pi_k^{(x)}.
\end{equation}
This means that \textit{any QP can be understood as a quantum controlled transformation at every point}. Indeed, every point $x$ is transformed to a new point $y_k(x)$, controlled by the orthogonal projections $\pi_k^{(x)}$. In the case where these all commute, the permutation is QC. 
That is, if all the $\pi_k^{(x)}$ commute, a common eigenbasis exists for every $k$ and $x$ and \eqref{eq:local_basis}
reduces to \eqref{eq:qcontrolled}. Namely, a set of $\pi'_k$ that is independent of $x$ exists so that
\begin{align}
    u_{QC} & = \sum_{x}^N \sum_{k=1}^{K} \ket{y_k(x)}\bra{x} \otimes \pi'_k \nonumber \\ & = \sum_k^K \sigma_k \otimes \pi'_k
\end{align}
where here $\sigma_k = \sum_x^N \ket{y_k(x)}\bra{x}$

In this sense, a BQC permutation is a \textit{local} superposition of transformations  (controlling over a different basis of $\mathcal{H}_O$ at every point) and a QC permutation is a \textit{global} superposition of transformations (controlling over one basis of $\mathcal{H}_O$ everywhere). We see concretely how this takes place in the first-quantisation formalism when the locations are graph nodes in Section \ref{sec:toygraph}, and more generally when the locations are positions on a line in Section \ref{sec:bqchamiltonian}. As remarked in Section \ref{sec:qft}, and as we will see in more detail in Sections \ref{sec:localizetwo} and \ref{sec:scalar}, the fact that BQC permutations are locally defined superpositions of transformations arises from the non-commutativity of reference fields.

\section{Quantum permutations as symmetries}\label{sec:toygraph}

We now see a first example of how `genuinely quantum' symmetries of graphs can yield new physical symmetries. For this, we use a toy model of a particle on a graph, which demonstrates intuitively the properties of BQC permutaitons discussed in Sections \ref{sec:bqc} and \ref{sec:localRelationality}. 

A prominent role of QPs in mathematics is that they introduce the notion of quantum symmetries of graphs, also known as quantum automorphisms. The definition of a quantum automorphism of a graph $\Gamma$ is a QP $u$ such that $u_{xy}u_{x'y'} = 0$ if $x \sim_\Gamma x'$ and $y \nsim_\Gamma y'$ (or vice versa). Here, $\sim_\Gamma$ signifies whether two nodes of the graph $\Gamma$ are connected by an edge. That is, recalling that $u_{xy}$ and $u_{x'y'}$ are orthogonal projections, $u$ preserves the graph structure if the subspace in which $x$ goes to $y$ is orthogonal to the subspace in which $x'$ goes to $y'$. This is equivalent to demanding that the adjacency matrix of the graph $A_\Gamma$ is invariant under $u$. That is, $u$ is a quantum automorphism of $\Gamma$  if and only if $A_\Gamma \otimes \mathbb{1}_O = u \; \big( A_\Gamma \otimes \mathbb{1}_O \big) \; u^\dagger$.

A classical automorphism (classical symmetry) of $\Gamma$ is a permutation $\sigma_k$ 
for which $A_\Gamma = \sigma_k A_\Gamma \sigma_k^\dagger$. If $\Gamma$  has more than one classical automorphisms, using \eqref{eq:qcontrolled} it is immediate to construct a QC permutation $u$ that is a quantum automorphism of $\Gamma$, namely  $u = \sum_{k=1}^K   \sigma_k \otimes \pi_k$, for any set of orthogonal projections $\pi_k$ that sum up to the identity. This is the reason that QC permutations are called `classical' permutations in the mathematical literature: they correspond to a quantum superposition of classical symmetries of graphs. Understood as transformations that leave observables (Hermitian operators) invariant, they are not `new' symmetries: a QC permutation built from classical automorphisms leaves invariant exactly the same observables as the underlying classical automorphisms do.\footnote{This is of course not to say that QC permutations act on \emph{states} the way classical permutations do. Entanglement, for instance, is preserved under any classical permutation but need not be preserved under a QC permutation \cite{cepollaroSumEntanglementSubsystem2025}.}

Quantum automorphisms that are BQC permutations are therefore a non-trivial extension of symmetries of graphs: strikingly, pairs of graphs can have a BQC permutation as its symmetry while it has \emph{none} classical symmetry. Therefore, \emph{BQC permutations allow to construct symmetries with no classical counterpart.} 

Imagine a quantum particle $\Hc_S$ and two reference systems $\Hc_{O_1}, \Hc_{O_2}\simeq\mathbb{C}^2$. The adjacency matrix of the graph can be thought of as an observable (Hermitian operator) given relative to $O_1$:
\begin{align}\label{eq:hamiltonian}
    A = \sum_{x\sim_\Gamma x'}\vert x \rangle \langle x' \vert \otimes \mathbb{1}_{O_1 O_2} = A_\Gamma \otimes \mathbb{1}_{O_1 O_2}.
\end{align}
Here, $\Gamma$ is a labeled graph with the node labels $x$ corresponding to some fixed basis $\ket x$ of $\Hc_S$, and we understand them here as positions relative to $O_1$ (similar to Section \ref{sec:1stq}). 

The general state of the particle can be written as $\vert\psi\rangle = \sum_{x}  \psi(x) \vert  x \rangle \otimes \vert \phi_{x} \rangle$ where $\psi(x)$ is the (discrete) wavefunction and $\ket{\phi_{x}} \in \Hc_{O_1} \otimes \Hc_{O_2}$ are states of the reference systems. The physical meaning of $A$ can be seen from its expectation value. Because the adjacency matrix is symmetric, for every edge between the nodes $i,j$ we get a term $\langle \phi_i \vert \phi_j \rangle$ + $\langle \phi_j \vert \phi_i \rangle$, and so
\begin{align}
    \langle \psi \vert A \vert \psi \rangle =\frac{1}{2} \sum_{i\sim_\Gamma j} \mathrm{Re} \langle \phi_i \vert \phi_j \rangle.
\end{align}
Then, the observable $A$ gives a measure of the alignment of the spin state across the edges on the graph. If the state and graph are such that each spin state pair connected by an edge is aligned (anti--aligned) on the two nodes then $\langle \psi \vert O \vert \psi \rangle$ is maximised (minimised), and it is zero if each pair of spins connected by an edge is orthogonal.  

Given a state $\vert\psi\rangle $, all states generated by acting on $\vert\psi\rangle $ with isomorphisms of the graph will give the same expectation value for the $A$. These are the classical symmetries of the graph $\Gamma$, classical permutations $\sigma$ on $\Hc_S$ that preserve the edge structure of $\Gamma$. For these, $\sigma^\dagger A_\Gamma \sigma = A_\Gamma$, and thus $\left( \sigma \otimes \mathbb{1}_{O_1 O_2} \right)^\dagger  A \left( \sigma \otimes \mathbb{1}_{O_1 O_2} \right) = A$. Quantum controlled magic unitaries $u = \sum_{k=1}^K   \sigma_k \otimes \pi_k$ with $\sigma_k$ being symmetries of $\Gamma$ also leave $A$ invariant, so that these magic unitaries extend its symmetry group.

BQC permutations that are symmetries of $\Gamma$ non-trivially extend the set of transformations leaving $A$ invariant. An example is the BQC permutation $u$ studied in Section \ref{sec:bqc} (see \eqref{eq:explbqc}, \eqref{eq:eg2} and Figure \ref{fig:decomp}):
\begin{align}
\label{eq:ugraph}
    u=
\begin{pmatrix}
\ket{\uparrow}\bra{\uparrow}_{O_2} & \ket{\downarrow}\bra{\downarrow}_{O_2} &  &  \\
\ket{\downarrow}\bra{\downarrow}_{O_2} & \ket{\uparrow}\bra{\uparrow}_{O_2} &  &  \\
 &  & \ket{+}\bra{+}_{O_2} & \ket{-}\bra{-}_{O_2} \\
 &  & \ket{-}\bra{-}_{O_2} & \ket{+}\bra{+}_{O_2} \\
\end{pmatrix}.
\end{align}
This is a BQC symmetry, for instance, of the 4-cycle (square) graph.

As discussed in Section \ref{sec:bqc}, this BQC permutation exchanges nodes 1 and 2 (3 and 4) on the down (plus) subspace and leaves them unchanged on the up (minus) subspace. Intuitively, the quantum permutation \eqref{eq:ugraph} quantum controls `simultaneously' on two non-commuting bases of the reference system $\Hc_{O_2}$ on different `parts' of the graph. This demonstrates the point discussed in Section \ref{sec:localRelationality}: the characteristic difference between the special case of QC and general case of BQC permutations is that the latter are constructed through non--commuting bases of the subspaces defined by different locations. In the above example, on the nodes $1,2$ the control basis is $\ket{\uparrow,\downarrow}$ and on the nodes $3,4$ it is the $\ket\pm$.

The toy-model studied in this Section used a first-quantisation formalism. In Section~\ref{sec:ising}, we will use the second-quantisation formalism to see how genuinely quantum symmetries of graphs are new symmetries for the Ising model. In the following two Sections, we turn to demonstrating that quantum permutations generate new symmetries also in physically relevant contexts where a graph structure is not present. 

\section{quantum controlled permutations as quantum reference frames} \label{sec:qrfs}

In this section we consider transformations developed in the context of the program of \textit{quantum reference frames (QRFs)}, e.g.~\cite{Zych:2015fka,Loveridge_2018,giacominiQRF2019,zychRelativityQuantumSuperpositions2018,hametteQuantumReferenceFrames2020,Vanrietvelde2020changeof,Castro-Ruiz_2020,Hardy:2019cef,krummQuantumReferenceFrame2021a,
AliAhmad_2022,giacominiQuantumSuperpositionSpacetimes2022,delahamettePerspectiveneutralApproachQuantum2021a,hohnTrinityRelationalQuantum2021,hohnEquivalenceApproachesRelational2021,delahametteQuantumReferenceFrames2023,Hoehn_2023,cepollaroSumEntanglementSubsystem2025,kabelQuantumCoordinatesLocalisation2025a,castro-ruizRelativeSubsystemsQuantum2025,delaHamette2022quantum,Chen_2026}, which is a main inspiration for this work. As the name suggests, this type of transformations is used to move between descriptions relative to different quantum systems. We will then see that the transformations that have been typically considered in the literature can be cast in the form of quantum controlled permutations. In the next Section, we see how to leverage the theory of quantum permutations to extend the quantum reference frame apparatus to beyond quantum controlled transformations. 

\begin{figure}
    \centering
    \hspace*{-2.5cm}
    \includegraphics[scale=1.2]{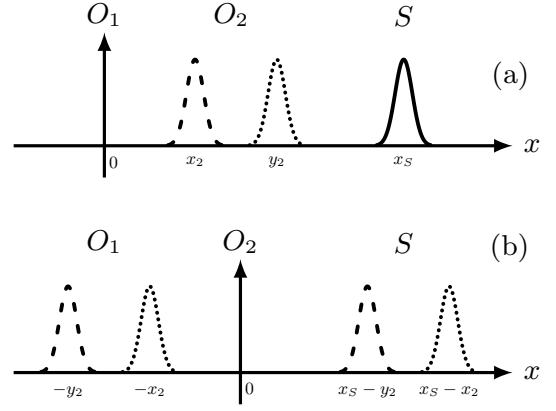}
    \caption{A simple use case for quantum controlled reference frame transformations. (a) A state from $O_1$’s perspective: $O_1$ is at the origin, $S$ is at a fixed position, and $O_2$ is in a superposition of two locations. (b) Same situation described from $O_2$’s perspective after applying the quantum reference frame transformation: $O_2$ is at the origin, while $O_1$ and $S$ are in an entangled state such that signed distances are preserved. Dashed and dotted lines represent different superposition branches.}
    \label{fig:qrfs}
\end{figure}

A QRF transformation is usually defined given a symmetry group $G$ and transforms the state of a tripartite system $\Hc_S \otimes \Hc_{O_1} \otimes \Hc_{O_2}$ from the frame of $O_1$ to that of $O_2$. The systems $O_1$ and $O_2$ are the observer/reference systems and $S$ is the system of interest. 
To briefly introduce QRFs, let us take the simple example when the group $G$ is the translations. Imagine a situation as depicted in Figure \ref{fig:qrfs}a. $O_1$ is in a definite location at $x=0$. From its perspective, $S$ is in a definite position $x=x_S$ and $O_2$ is in a superposition of locations $x_2$ and $y_2$. The state is
\begin{align}
\label{eq:state1}
    \ket\psi^{O_1} = \ket{x_S}_S \otimes \ket 0_{O_1} \otimes   \frac{1}{\sqrt 2} \Big( \ket{x_2} + \ket{y_2} \Big)_{O_2}.
\end{align}
Quantum reference frame transformations allow to `jump' to the perspective of $O_2$, even though $O_2$ is in a quantum state (in a superposition of locations). That is, a QRF transformation changes the description so that the location of $O_2$ becomes definite (say at $x=0$) while the (signed) distances, a relational observable, is conserved. This encodes the presence of translational symmetry. The only state that preserves the distances while $O_2$ is at $x=0$ is
\begin{align}
\label{eq:state2}
    \ket\psi^{O_2} =
    &\frac{1}{\sqrt 2} \Big( \ket{x_S - x_2}_S \ket{-x_2}_{O_1} + \ket{x_S-y_2}_S \ket{-y_2}_{O_1} \Big) \\
    &\otimes  \ket 0_{O_2} \nonumber
\end{align}
which is depicted in Fig.~\ref{fig:qrfs}b. By inspection, we see that the transformation which sends $\ket\psi^{O_2} \rightarrow \ket\psi^{O_1}$ will control on each of the superposed locations of $O_2$, and perform the required translation to the other systems accordingly, in such a way as to preserve the distances.

This is called a QRF transformation, and can be written in general as \cite{delahametteQuantumReferenceFrames2023}
\begin{align}
\label{eq:QRF}
 S^G_{O_1\rightarrow O_2}
=
 \text{SWAP}_{O_1O_2}
\;\;
\sum_{g\in G}  T^\dagger_{g, S} \otimes  \ket{ g^{-1} } \bra{ g }_{O_2},
\end{align}
where $\ket{g}:=T_g \ket{e}$ for all $g\in G$ form an orthonormal basis constructed by acting with $G$ on a fixed state $\ket{e}$ with the left regular representation $T_g$.\footnote{We leave implicit that $T_g$ are formally different operators on each Hilbert space, i.e.~we omit writing $T_g^{O_1}$, $T_g^{O_2}$ and $T_g^S$.} In the example discussed above, taking
 $G$ to be the translation group the QRF transformation \eqref{eq:QRF} sends the state \eqref{eq:state1} to the state \eqref{eq:state2}. It is important to notice that every $T_g$ is a \emph{permutation} of the basis elements $\ket{g}$: $T_g \ket{g'} = \ket{gg'}$, that is, $T_g$ takes basis elements to basis elements and is invertible. The operator $\text{SWAP}_{O_1O_2}$ exchanges the $O_1$ and $O_2$ elements of the tensor product.\footnote{Note that while it is common to define QRF transformations by fixing a symmetry group $G$, this is not necessary for the usual formal definition. It is sufficient to fix a basis $\ket{g}_S$ for $\Hc_S$, and bases $\ket{g}_{O_1,O_2}$ to control over. The $T_g$ can be taken to be unitaries on $\Hc_S$ that permute the basis elements $\ket{g}_S$. The additional flexibility here is because quantum permutations are formed by choosing a \emph{different basis} at each point, see \eqref{eq:local_basis}.}

The QRF transformation \eqref{eq:QRF} can be cast in the form of a QC permutation (up to a unitary on the observers' sector), since it can be written as 
\begin{align}\label{eq:qrfqc}
S^G_{O_1\rightarrow O_2}
=
\left( \mathbb{1} \otimes \tilde V \right) \;\;
\sum_{g\in G}  T^\dagger_g \otimes  \ket{ g } \bra{ g }_{O_2}.
\end{align}
Here, $\tilde V  = \text{SWAP}_{O_1O_2} \;\; \sum_g \ket{g^{-1}}\bra{g}_{O_2}$ is a unitary only on the observers' sector $O_1$ and $O_2$.

Given a Hamiltonian, any quantum permutation that is a superposition of permutations that are symmetries of the Hamiltonian is itself a symmetry. For example, in a $P+2$ partite system $S_1,S_2...S_P,O_1,O_2$ any central force Hamiltonian of the form
\begin{align}
    H = \sum_i \frac{\hat{p}_{S_i}^2}{2m_{S_i}} + \sum_{i\neq j} V\left(\left|\hat{x}_{S_i} - \hat{x}_{S_j}\right|\right)
\end{align}
with $m_{S_i}$ the mass of particle $S_i$ retains its form under translational QRF transformations \cite{delahametteQuantumReferenceFrames2023}. It is therefore invariant under the corresponding QC permutation of the form \eqref{eq:qrfqc}.\footnote{Note that for the case of reference frames obeying some dynamics themselves, $H$ in one frame should also contain a term describing the other (e.g.\ $\hat p_{O_i}^2/2m_{O_i}$), which would violate its invariance. Invariance would then be recovered in the `classical limit' of large masses of the reference frames.}

We have seen that QC permutations, understood as QRF transformations, can be symmetries of the dynamics of a physical system. We now proceed to give examples of `genuinely quantum' reference frame transformations, using BQC permutations. 

\section{`Genuinely quantum' reference frames}\label{sec:qrfsplus}

In the previous Section we saw that typical QRF transformations studied in the literature can be cast in the form of quantum controlled permutations. However, as discussed toward the end of Section \ref{sec:qft}, the non-commutativity of quantum theory suggests that generic transformations between quantum systems of references will not be of the quantum controlled form. We will now work out two examples of BQC permutations used as a `genuinely quantum' reference frame transformation: (A) in first-quantization formalism, demonstrating they can be used to control on non--commuting observables at different locations, and (B) in second-quantization formalism, demonstrating that they can localize states that can not be localized with QC transformations.

\subsection{BQC frames in quantum mechanics}
\label{sec:bqchamiltonian}
Recall the example discussed in Section ~\ref{sec:1stq}, of a particle on a line. The relative observable was the (signed) distance $\hat X_1$ ($\hat X_2$) as seen by $O_1$ ($O_2$). The QP that changes from the frame of $O_1$ to the frame of $O_2$ is
\begin{align}
    u &= \sum_{x} T_{-x, S} \otimes \ket{x}\bra{x}_{O_2}  \\\nonumber
    \hat X_2 &= \hat X_1 - \hat X_{O_2} = u \; \hat X_1 \; u^\dagger,
\end{align}
with $T_{-x}\ket{y}_S = \ket{y-x}_S$ translating the $\hat X_1$ eigenbasis. 

In order for $\hat X_2$ to make sense as a relative to $O_2$ observable, we had assumed it  only acts non-trivially on the state of $O_2$. More precisely, we had assumed that for every $x_1$ there exists an observable $\hat X_{2,x_1}$ such that $\hat X_2\ket {x_1} \ket\phi = \ket {x_1} \otimes \hat X_{2,x_1} \ket\phi$ for all $\ket\phi\in\Hc_{O_2}$

The difference between QC and BQC transformations is directly linked to the commutativity between the operators $\hat X_{2,x_1}$ for all $x_1$. When they all commute, their projections onto their eigenspaces $u_{x_1x_2}$ will also all commute, and the QP that results will be QC. When the  $\hat X_{2,x_1}$ do not all commute, it will be BQC.  In the example worked out in Section \ref{sec:1stq} and recalled above, the $\hat X_{2,x_1}$  commute for different $x_1$, since $\hat X_{2,x_1} = x_1 \mathbb 1_{O_2} - \hat X_{O_2} $. An example of a relative observable arrived at by a BQC permutation is when the observer's position is used to transform \emph{some} $\hat X_1$ (eigen)values, while the observer's \emph{momentum} is used for other (eigen)values of $\hat X_1$. Let us now work this out explicitly.

\subsubsection*{Controlling on two complementary variables}

First, recall that the typical QRF transformations found in the literature are quantum controlled (see Section \ref{sec:qrfs}). In particular, this includes the case when a symmetry is assumed corresponding to a group that is the direct product of two or more other groups, and of which the corresponding observables commute. In such cases, the corresponding Hilbert space can be broken down to a direct product of two spaces, each corresponding to a distinct symmetry: the resulting QRF transformation effectively acts on each space separately. 

Our task here is to construct, instead, \emph{a QRF transformation which quantum controls on two subspaces corresponding to non-commuting observables, such as momentum and position}.  In this case, the Hilbert space cannot be broken down to a product corresponding to the different symmetries, as the corresponding relational observables will not commute, and it will correspond to a BQC permutation.

This is fundamentally different from the case of a non-ideal quantum reference frame. The latter could also control on two non-commuting observables --- however only by virtue of the reference states being no longer perfectly distinguishable, i.e.\ the infinite resource providing perfect precision to the frame is given up. Here, by contrast, the reference states \textit{are} perfectly distinguishable ($u_{x_2x_1}u_{x_2x_1'}=0$ for $x_1\neq x_1'$ and $u_{x_2x_1}u_{x_2'x_1}=0$ for $x_2\neq x_2'$), and what is given up instead is the requirement that the same control basis be used at every point.

Accordingly, we will say that two groups $G_1,G_2$ are \emph{incompatible} if the orthonormal bases created by their (left regular, see previous Section) representations $ \{ \ket{g_1} \}_{g_1 \in G_1}$ and $ \{ \ket{g_2} \}_{g_2 \in G_2}$  \emph{do not commute}. That is, when at least for one pair $g_1, g_2$ it holds that $\ket{g_1}\bra{g_1}$ and $\ket{g_2}\bra{g_2}$ do not commute.

For any single group $G$, all the projectors $\ket{g}\bra{g}$ commute by construction. Thus, \emph{it is not possible to control on two incompatible groups with an ideal QC permutation.} The non--commuting projections will directly enter the definition of the quantum permutation which encodes the transformation. Since non--commuting entries are present, \emph{in order to control on two incompatible groups, the only option is to quantum control on subspaces of the system of interest through a BQC permutation.}

Consider that we have two incompatible symmetry groups $G_1, G_2$ e.g.~translations in position and boosts in momentum. A BQC permutation that is block diagonal --- of the general form \eqref{eq:bqcblocks} --- could be written as
\begin{align}\label{eq:twog}
u=
\sum_{g_1 \in G_1} u_{g_1} \otimes \ket{g_1}\bra{g_1}
+
\sum_{g_2 \in G_2} v_{g_2} \otimes \ket{g_2}\bra{g_2},
\end{align}
where $u_{g_1} v_{g_2}^\dagger = 0$, $u_{g_1} u_{g_1}^\dagger = I_1$,  $v_{g_2} v_{g_2}^\dagger = I_2$, for all $g_1 \in G_1, g_2 \in G_2$, with some operators $I_1 +I_2= \mathbb{1}$. Note that the above equation is in the form of \eqref{eq:bqcblocks} with two blocks, corresponding to the two groups.

Let us now take that $O_1, O_2$ and $S$ are particles on a line with position and momentum operator. We denote $\hat X_{O_2}, \hat X_S$ and $\hat P_{O_2}, \hat P_S$ the position and momentum of $O_2$ and $S$ respectively relative to $O_1$. Assume that $S$ and $O_2$ interact through a Hamiltonian that depends on the absolute ratio of their positions if $S$ is positioned at or to the right of $x=0$, while it depends on the absolute product of $S$'s position and $O_2$'s momentum otherwise:
\begin{align}\label{eq:qrfham}
   H = \pi_{\hat X_S\geq 0} V \left( x_0 \left| \frac{\hat X_S }{ \hat X_{O_2}} \right|  \right) + \pi_{\hat X_S < 0} V \left( -\frac{x_0}{\hbar} |\hat X_S  \hat P_{O_2} |  \right),
\end{align}
with $\pi_C$ a projector on all states that fulfill the condition $C$. Then, we can write a quantum permutation $u$ that transforms this state from $O_1$'s perspective to $O_2$'s perspective:
\begin{align}
    u =
    &\sum_{x}  u_{x} \otimes \ket{x}\bra{x}
   +\sum_{p}  v_{p} \otimes \ket{p}\bra{p},
   \\\nonumber
    u_{x} =&\; \sum_{x_S \geq 0} \ket{ \; x_0 \left| x_S / x \right| \;}\bra{x_S},
   \\\nonumber
   v_{p} =&\; \sum_{x_S < 0} \ket{ \; -x_0 |x_S  p | /\hbar \;}\bra{x_S},
\end{align}
This is a BQC QRF transformation that depends on the non-commuting observables $\hat X_{O_2}, \hat P_{O_2}$. It is a symmetry of Hamiltonians of the form \eqref{eq:qrfham}. Note that above we left implicit the state of $O_2$, which after the transformation is fixed to $\ket{x_{O_2}=x_0}$ for $x_S\geq 0$ and to $\ket{p_{O_2} = \hbar/x_0}$ for $x_S < 0$. Indeed, applying the above $u$ on a basis we get
\begin{align}\label{eq:qrfbqc}
    u \ket{x_S}\ket{x} &= \ket{\; x_0 \left| x_S / x \right|\;}\ket{x}  &\text{if } x_S\geq 0\\\nonumber
    u \ket{x_S}\ket{p} &= \ket{\; -x_0 |x_S  p | / \hbar \; }\ket{p} &\text{if } x_S < 0
    .
\end{align}
Here, we see an example of the fact that BQC permutations allow to control on subspaces defined by different locations using arbitrary choices of bases (see
\eqref{eq:local_basis} and surrounding discussion). The above transformation uses the momentum and position basis, which do not commute, on different parts of the real line.

\subsection{BQC frames in quantum fields}
\label{sec:localizetwo}

In Section \ref{sec:qft}, we saw that the QP \eqref{eq:qpfield} transforming between the two reference systems was of the form
\begin{align}
    u_{x_2x_1} = \mathbb1_S\otimes\sum_{q} P_1^{qx_1}  \otimes  P_2^{qx_2} \nonumber.
\end{align}
where the $P_i^{qx_i}$ are orthogonal projectors on the eigenspace of the reference field $\hat{\chi}_i(q)$ with eigenvalue $x^\mu$ (it assigns the $x^\mu$ coordinate to the abstract point $q$).
We remarked that this QP will generally be BQC, since we generally expect that for quantum fields there exist some $q,q'$ such that $[ \hat\chi_i(q), \hat\chi_i(q') ] \neq 0$ (known as the principle of microcausality). This implies that $[P_i^{qx_i}, P_i^{q'x_i'}] \neq 0$ for some $x_i,x_i'$ and therefore $[ u_{x_2 x_1}, u_{x_2' x_1'}] \neq 0$. Strikingly, this implies that the non-commutativity of fields in quantum field theory is therefore directly linked to `genuinely quantum' reference frames. Let us now work out an example of what can be achieved with a BQC QRF and which is not possible with a QC QRF.

\subsubsection*{Localizing two particles simultaneously}

We now use the discrete quantum fields of Section \ref{sec:qft} to give an example of a BQC permutation that transforms a bipartite state in which both particles are initially in indefinite states, to a state frame in which both particles become definite. We then show that it is impossible to transform this state to a quantum reference frame in which both particles become definite using a quantum controlled transformation. 

Assume two massive particles $A,B$ described by quantum scalar fields $\hphi_A, \hphi_B$ with Fock spaces $\Hc_A,\Hc_B$ and two reference fields $\hat\chi_1, \hat\chi_2$ with Fock spaces $\Hc_{O_1},\Hc_{O_2}$. The total space is then $\Hc_A \otimes \Hc_B \otimes \Hc_{O_1}\otimes\Hc_{O_2}$. We assume the particles $A$ and $B$ to be masses with non-relativistic motion, in the sense that $mc\Delta x \gg \hbar$, where $\Delta x$ is the scale of difference between eigenvalues of the reference fields $\hat\chi_{1,2}$. The fields $\hat\chi_1, \hat\chi_2$ satisfy the conditions set out in Section \ref{sec:qft} in order to be good reference fields, and are otherwise arbitrary.

Assume further that the reference field $O_1$ is definite. This means that there is exactly one value $x_q\in\mathbb{R}^4$ for every $q\in\Mc$: $\hat\chi_1(q) = x_q \mathbb{1}_{O_1}$. In terms of the projectors, $P_1^{qx} = \delta_{xx_q} \mathbb{1}_{O_1} $.

Now, consider the state $\ket\psi\in \Hc_A \otimes \Hc_B \otimes \Hc_{O_1}\otimes\Hc_{O_2}$
\begin{align}
\label{eq:toLocalize}
\ket\psi = 
    &\frac{1}{\sqrt2}\ket{x_A}^{O_1} \frac{\ket{x_B}^{O_1} +  \ket{x_B'}^{O_1}}{\sqrt2} \ket{\phi}_{O_1 O_2}
    \;+
    \\\nonumber
    &\frac{1}{\sqrt2}\ket{x_A'}^{O_1} \frac{\ket{x_B}^{O_1} - \ket{x_B'}^{O_1}}{\sqrt2} \ket{\phi^\perp}_{O_1 O_2}.
\end{align}
Here, the $x_A,x_A',x_B,x_B'\in\mathbb{R}^4$ are all different, $\ket\phi,\ket{\phi^\perp} \in \Hc_{O_1} \otimes \Hc_{O_2}$ are any two orthogonal states of the joint system of reference fields and $\ket{\phi^\pm} = \frac{1}{\sqrt2} \left(  \ket\phi \pm \ket{\phi^{\perp}} \right)$.\footnote{Note that the appearance of two orthogonal reference states in \eqref{eq:toLocalize} is consistent with $O_1$ being definite: $\ket\phi$ and $\ket{\phi^\perp}$ correspond to one and the same definite assignment of coordinates by $O_1$, all of the indefiniteness resides in $O_2$.}
A state $\ket{y}^{O_1}$ describes a particle being created at the specified spacetime point relative to $O_1$:
\begin{align}
    \ket{y_A}^{O_1}:&= a_1^\dagger(y_A)\ket{\Omega_A} = a^\dagger(q_A)\ket{\Omega_A} 
    \\\nonumber
    \ket{y_B}^{O_1}:&= b_1^\dagger(y_B)\ket{\Omega_B} = b^\dagger(q_B)\ket{\Omega_B} 
\end{align}
for any $y_{A.B},q_{A,B}$ such that $P_1^{q_Ay_A} = P_1^{q_By_B} = \mathbb{1}_{O_1}$. The states $\ket{\Omega_A}\in\Hc_A, \ket{\Omega_B}\in\Hc_B$ are the vacuum states of the $A$ and $B$ fields. The annihilation operators for $A,B$ were denoted $a,b$ and a subscript $1$ denotes the field relative to $O_1$ as defined in \eqref{eq:phi_i}.

We now ask whether there exists a transformation, for which relative to $O_2$ the state is
\begin{align}
\label{eq:o2perspective}
    \ket\psi = 
    a_2^\dagger(x_A) b_2^\dagger(x_B)  \;\ket\Omega_A \ket\Omega_B \ket{\phi}_{O_1 O_2}.
\end{align}
That is, so that \emph{both} particles are localised. We show in Appendix \ref{ap:localizetwo} that this transformation from $O_1$ to an $O_2$ indeed exists. Using the notation of Section \ref{sec:qft}, in particular compare to \eqref{eq:qpfield}, the BQC quantum reference frame which achieves the double localization is given as follows
\begin{align}
u_{x_A x_A} &= \mathbb{1}_{AB}\otimes\sum_q P_1^{q x_A}\otimes P_2^{q x_{A}} = \ket{\phi}\bra{\phi} \nonumber \\
u_{x_A x_A'} &= \mathbb{1}_{AB}\otimes\sum_q P_1^{q x_A}\otimes P_2^{q x_{A}'} = \ket{\phi^\perp}\bra{\phi^\perp} \nonumber  \\
u_{x_A' x_{A}} &= \mathbb{1}_{AB}\otimes\sum_q P_1^{q x_A'}\otimes P_2^{q x_{A}} = \ket{\phi^\perp}\bra{\phi^\perp} \nonumber  \\
u_{x_A' x_{A}'} &= \mathbb{1}_{AB}\otimes\sum_q P_1^{q x_A'}\otimes P_2^{q x_{A}'} = \ket{\phi}\bra{\phi} \nonumber \\
u_{x_B x_{B}} &= \mathbb{1}_{AB}\otimes\sum_q P_1^{q x_B}\otimes P_2^{q x_{B}} = \ket{\phi^+}\bra{\phi^+} \nonumber \\
u_{x_B x_{B}'} &= \mathbb{1}_{AB}\otimes\sum_q P_1^{q x_B}\otimes P_2^{q x_{B}'} = \ket{\phi^-}\bra{\phi^-} \nonumber \\
u_{x_B' x_{B}} &= \mathbb{1}_{AB}\otimes\sum_q P_1^{q x_B'}\otimes P_2^{q x_{B}} = \ket{\phi^-}\bra{\phi^-} \nonumber \\
u_{x_B' x_{B}'} &= \mathbb{1}_{AB}\otimes\sum_q P_1^{q x_B'}\otimes P_2^{q x_{B}'} = \ket{\phi^+}\bra{\phi^+}.
\end{align}
 Note that this is the quantum permutation \eqref{eq:eg2} with $\pi_1 = \ket\phi\bra\phi, \pi_2=\ket{\phi^+}\bra{\phi^+}$. $P_i^{qx}$ is the projection on the $x$ eigenspace of $\chi_i(q)$.

In the resulting quantum coordinate system (as defined by the reference test field $\hat{\chi}_2$), both particles are created in definite spacetime points. This is not possible to achieve with a quantum controlled permutation.\footnote{This statement is made with regard to the specific state \eqref{eq:toLocalize}. For different initial states of the system, or for systems with an additional degree of freedom that helps distinguish the branches and can be controlled over, QC transformations may very well be able to transform away superpositions \cite{kabelQuantumCoordinatesLocalisation2025a,delaHamette2022quantum}.} The proof, given in Appendix~\ref{ap:localizetwo}, is a formalization of the following observations. The state $\ket\psi$ is not a superposition of definite bipartite states: for $\ket{\tilde\phi}\in \Hc_{O_1}\otimes \Hc_{O_2}$, there does not exist a state $\braket{\tilde\phi\vert\psi}\in\Hc_A\otimes\Hc_B$ that has both particles in definite states. In other words, $\braket{\tilde\phi\vert\psi}  \neq \ket{\tilde x_A}^{O_1} \ket{\tilde x_B}^{O_1} $ for any $\tilde x_A, \tilde x_B$. Therefore, a QC permutation, which is a superposition of definite permutations, cannot render the state definite.

Let us now comment on how the above are linked to the observation made in Section \ref{sec:localRelationality} that BQC permutations can be understood as making a local choice of control basis for the observers sector $\mathcal{H}_{O_1} \otimes \mathcal{H}_{O_2}$ at each point. We had seen that quantum permutations can be written in the form \eqref{eq:local_basis}, repeated here for convenience
\begin{equation}
    u = \sum_{x}^N \sum_{k=1}^{K_x} \ket{y_k(x)}\bra{x} \otimes \pi_k^{(x)}. \nonumber
\end{equation}
In the example we studied above, the new reference field $\hat{\chi}_2$ does not commute with itself at the points $q_A$ and $q_B$ (for instance), which correspond to the coordinates $x_A$ and $x_B$ relative to $O_1$ (assigned by the $\hat{\chi}_1$ reference field). That is, 
$[\hat\chi_2(q_A),\hat\chi_2(q_B)] \neq 0$. 

The above example demonstrates that there exist states (such as \eqref{eq:toLocalize}), which in order to localize by an appropriate choice of frame, the reference field will not commute with itself at some points. \emph{Therefore, the fact that quantum fields do not commute at different points compels the extension of the QRF framework to include BQC transformations. }

\begin{figure}
\hspace{-1.5cm}
\includegraphics[width=1\linewidth]{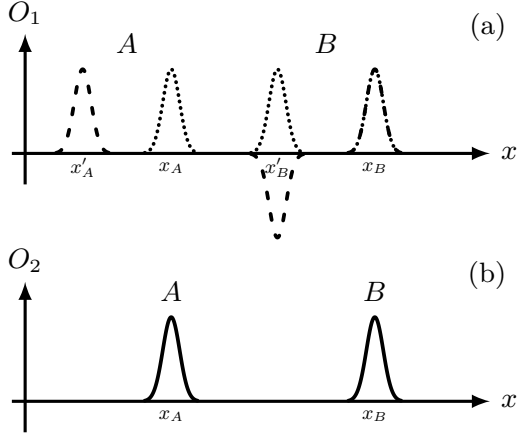} 
    \caption{(a) The state \eqref{eq:toLocalize} of the two particles with respect to observer $O_1$ is a superposition of two terms: a term where $A$ is localized at $x_A$ and $B$ is in a superposition of locations $x_B$ and $x_B'$, and a term where $A$ is localized at $x_A'$ and $B$ is in a superposition of the same two locations but now with a relative minus sign. Here, the different line patterns represent the entanglement across branches. In the $O_1$ frame the particles are also entangled with the reference fields, which are in the states $\ket\phi_{O_1O_2}$ and $\ket{\phi^\perp}_{O_1O_2}$ in the respective branches. These field states are not shown in the figure. (b) From the perspective of $O_2$, the same state is \eqref{eq:o2perspective}, with $A$ and $B$ at the definite locations $x_A$ and $x_B$. In contrast to the frame of $O_1$, in the frame of $O_2$ there is no entanglement, neither among the particles nor between the particles and the fields.}
    \label{fig:localizetwo}
\end{figure}

\subsection{Brief comments on relevant QRF literature}\label{sec:literature}
We pause to make some comments concerning some particularly relevant QRF literature.

A QRF transformation that achieves a similar goal to that presented in Section \ref{sec:bqchamiltonian} appeared in \cite{castro-ruizRelativeSubsystemsQuantum2025}. 
 In that work, an additional degree of freedom was introduced in order to control on the bases $\ket x$ and $\ket p$. In the notation of that work it is a variable $\theta$. A basis of the joint system is constructed of the form $\{ \ket x\ket\theta \}_{\theta\in\Theta} \cup \{ \ket p \ket\theta \}_{\theta\notin\Theta}$, with $\Theta$ some subset of possible $\theta$ values. This allows to use a QC transformation in order to control on both $\ket x$ and $\ket p$. This is different than what we have done, as we did not introduce a new degree of freedom. Rather, we gave up on the restriction to QC transformations. 

It is also interesting to briefly discuss the case of non-ideal QRFs \cite{hohnEquivalenceApproachesRelational2021,hohnTrinityRelationalQuantum2021,delahamettePerspectiveneutralApproachQuantum2021a}. The QRF transformations we have discussed in this manuscript, are known as ideal QRFs. This is because it is assumed that different states of an observer (corresponding to different group elements) are perfectly distinguishable (orthogonal). Non-ideal QRFs relax this restriction, and by virtue of that may allow for simultaneous control of non-commuting observables (although, we are not aware of a work demonstrating this). The interpretation of quantum permutations that we have presented is as ideal quantum reference frames. That is, we have assumed that any two $O_2$ states that disagree on the new value $y\neq y'$ of a point $x$ relative to $O_1$ are orthogonal, i.e.\! $u_{xy}u_{xy'}=0$. The fact that they allow simultaneous control on complementary variables is not due to the introduction of an uncertainty threshold for the observer's state (which is the characteristic of non-ideal QRFs), but rather in not requiring that the basis for $\Hc_O$ is chosen globally. As we have seen, BQC permutations are constructed by making an arbitrary choice of basis for $\Hc_O$ at each location.

Turning to the second example presented in Section \ref{sec:localizetwo}, this was inspired by the work \cite{kabelQuantumCoordinatesLocalisation2025a}, which used QC transformations to define quantum coordinate changes defined through reference fields. Section \ref{sec:localizetwo} (and Section \ref{sec:qft}) extend what was done in \cite{kabelQuantumCoordinatesLocalisation2025a} to incorporate the fact that quantum fields do not commute at different points. Because BQC permutations allow the control basis for the transformation to be different at every point, the transformation is not definite or `classical-like' for any post-selection on $\Hc_O$ (meaning, after projecting on an observer's state). This is precisely what allows the transformation to not preserve `relative definiteness', namely localizing $B$ while not delocalizing $A$, even after post selection on $\ket\phi$ or $\ket{\phi^\perp}$ (see end of Section \ref{sec:localizetwo}).

Let us see why a QC permutation, such as those discussed in \cite{kabelQuantumCoordinatesLocalisation2025a}, does not permit for localization of $A$ and $B$. In simplified notation, the QC transformation would act on some state $\sum_\chi \sigma_\chi \otimes \ket{\chi}\langle \chi | \Psi \rangle$, with $\sigma_\chi$ describing a definite transformation of $A,B$ and  $\ket\chi$ the observer states. Post-selecting on a basis element $\ket{\chi_*}$, the picture becomes definite or classical-like, that is,  $\ket{\chi_*}\bra{\chi_*} \sum_\chi \sigma_\chi \otimes \ket{\chi}\langle \chi | \Psi \rangle =\sigma_{\chi_*} \otimes \ket{\chi_*}\langle \chi_* | \Psi \rangle $. Therefore, for every state $\ket{\chi_*}$ of the observer, the relative definiteness of $A,B$ cannot be changed, as classical permutations take definite locations to definite locations and superpositions to superpositions. Intuitively, because BQC permutations do not restrict to a branch-by-branch classicality of the transformation, they are able to change the relative definiteness.

Finally, a recent work formulates quantum reference frames for linearized quantum gravity by introducing quantum reference fields \cite{Chen_2026}. The transformations obtained in that work are quantum controlled unitaries --- the control being the relative displacement between the two reference fields. This is consistent with their restriction to spacelike hypersurfaces, on which the fields' commutation relations are trivial. Therefore, it is interesting to investigate what would be their representation as QC permutations.

In summary, beyond quantum controlled permutations introduce a dependence of the control basis on $x$ values, which allows for more general interplay between the quantum coordinate system and the quantum objects which it describes.

\section{genuinely quantum symmetries of the Ising model}\label{sec:ising}
In this section we will see how the formalism we have developed, whereby QPs are understood as quantum reference system transformations, can be used to demonstrate new symmetries of the Ising model.

The Ising model is a system of spins on the nodes of a graph $\Gamma$ such that only neighboring spins directly interact. In addition, an external magnetic field affects all spins. The Hamiltonian is
\begin{align}
\label{eq:IsingH}
    H = - J \sum_{{x} \sim_\Gamma  {x}'} \hat \sigma_1({x}) \hat \sigma_1({x}') - h \sum_{{x}} \hat \sigma_1({x}),
\end{align}
with constants $J,h\geq 0$ representing the strength of interaction and the magnetic field respectively. The operator $\hat \sigma_1({x}) =\ket\uparrow\bra\uparrow_{1, x}-\ket\downarrow\bra\downarrow_{1, x}$ is the spin operator on $\Hc_{1, {x}}$, which represents the spin on the node of the graph that is at ${x} \in \mathbb{R}^n$. The state space is $\Hc_S \simeq \bigotimes_{{x}} \Hc_{1,{x}}$, with an orthonormal basis $\ket{s} = \bigotimes_{{x}} \ket{s_{ x}}$, $s_{{x}} = \uparrow,\downarrow$.

Below, we combine the the second quantization formalism presented in Section \ref{sec:qft} and \ref{sec:localizetwo}, and the first quantization toy-model of a spin on a graph of Section \ref{sec:toygraph}. We interpret ${x}$ to be labels of the nodes of the graph,  given according to some reference. To transform $\sigma_1({x})$ to the reference frame of $O_2$, we simply write 
\begin{align}
\label{eq:QPonscalar}
   \hat \sigma_2( y) = \sum_{{x}} \hat \sigma_1({x})  {u_{ y  x}}
\end{align}
as was found in Section \ref{sec:qft}, see \eqref{eq:phi12rule}. The Hamiltonian from $O_2$'s reference then looks like:
\begin{align}
    H = - J \sum_{ y ,  y'}  \hat \sigma_2( y) A_{2, y y'} \hat \sigma_2( y')   - h \sum_{ y} \hat \sigma_2( y),
\end{align}
with $A_{2, y  y'} = \sum_{{x}, {x}' }A_{1, x x'} u_{ x y}u_{ x' y'} = (u^\dagger A_1 u)_{ y y'}$ and $A_1$ the adjacency matrix of the graph, that is $A_{1, x x'}=1$ if ${x} \sim_\Gamma  {x}'$ and else $A_{1, x x'}=0$. Symmetries of the Ising model are reference system transformations $u$ such that the Hamiltonian retains its form -- or in other words, such that the equations of motion do not change. This only happens if $u\,A_1\,u^\dagger = A_1$, which is exactly the definition of quantum automorphisms of the graph.

The usual symmetries of the Ising model are classical permutations $u$ that are automorphisms of $\Gamma$. As discussed in Section \ref{sec:toygraph}, if the graph has more than one classical symmetries, QC permutations that are quantum superpositions of those will also be symmetries of the Ising model. If the graph has genuinely quantum symmetries, this corresponds to the existence of BQC permutations that are quantum automorphisms of $\Gamma$.

\begin{figure}
    \centering
    \vspace{0.8cm}
    \includegraphics[width=\linewidth]{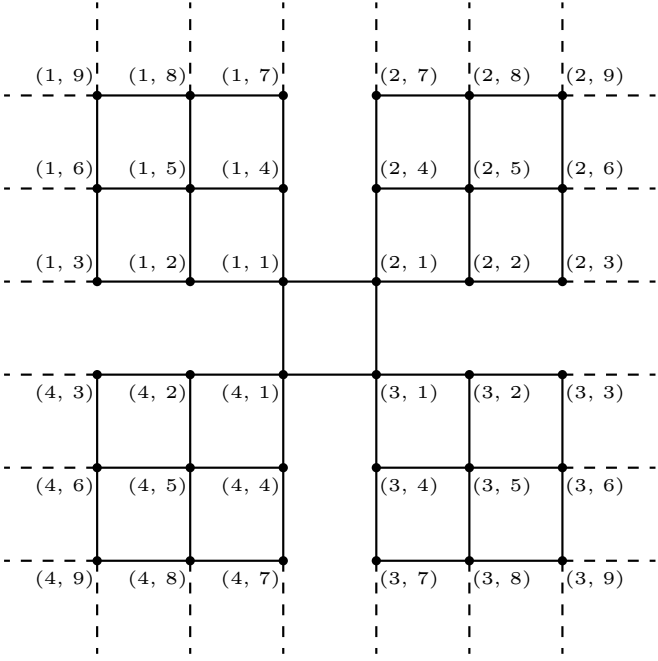}
    \caption{An example of a `large' graph whose corresponding Ising model is symmetrical under the quantum reference system transformation \eqref{eq:isingsymm}. This was constructed by extending the genuinely quantum symmetry of the 4-cycle discussed in Section \ref{sec:toygraph}. As far as we are aware, it is not known whether regular lattices (such as a square lattices) possess genuinely quantum symmetries, as this is not trivial to check.}
    \label{fig:ising}
\end{figure}

Let us see a concrete example. Consider the case of the graph $\Gamma$ depicted in Fig.~\ref{fig:ising}. A quantum symmetry for the Ising model on this graph is achieved by moving to some other reference $O_2$ such that the corresponding $u$ from \eqref{eq:qpfield} is a quantum symmetry of this graph. An example for such a $u$ is
\begin{align}\label{eq:isingsymm}
    u_{ y x} = 
    \begin{cases}
         \pi_1 &  x =  y = (1,j) \text{ or } \\& x =  y = (3,j)\\
         \mathbb{1} - \pi_1 &  x = (1,j),  y = (3,j) \text{ or }\\&  y = (1,j),  x = (3,j)\\
         \pi_2 &  x =  y = (2,j) \text{ or }\\&  x =  y = (4,j)\\
         \mathbb{1} - \pi_2 &  x = (2,j),  y = (4,j) \text{ or }\\&  y = (2,j),  x = (4,j) \\
         0 & \text{else}
    \end{cases}
\end{align}
with $\pi_1,\pi_2$ two non-commuting orthogonal projections on $\Hc_{O_1} \otimes \Hc_{O_2}$, which can be constructed by taking any two orthogonal states in that space $\braket{\phi  \vert \phi^\perp}=0$ and setting $\ket\pm = \frac{1}{\sqrt2} \left(  \ket\phi \pm \ket{\phi^{\perp}} \right)$, $ \pi_1 = \ket\phi\bra\phi, \pi_2=\ket+\bra+$.

Note that the same holds if one ``closes" the graph by adding the the edges $(1,n^2)\leftrightarrow(2,n^2)$, $(2,n^2)\leftrightarrow(3,n^2)$, $(3,n^2)\leftrightarrow(4,n^2)$, $(4,n^2)\leftrightarrow(1,n^2)$, with $n^2$ being the number of nodes at every quarter of the graph (9 of which are drawn in Fig.~\ref{fig:ising}). This would make the model admit cyclic boundary conditions.

In the above example, we used a rather ad-hoc large graph, which we constructed by extending the genuinely quantum symmetry of the 4-cycle discussed in Section \ref{sec:toygraph}, see  Figure \ref{fig:ising}. It should be noted that
there are theorems that show that almost all general graphs do not have quantum symmetries \cite{LupiniMancinskaRoberson2020}, while almost all tress do \cite{Junk_2020}. This is  situation is similar as for classical symmetries \cite{ErdosRenyi1963}. However, as far as we are aware, it is not known whether regular lattices possess genuinely quantum symmetries, and this is difficult to check for any given large graph. On the other hand, trees typically arise when causal structure is present.

On this note, it is interesting to discuss what is the physical interpretation of the reference frame that corresponds to genuinely quantum graph symmetry, according to the results in the preceding Sections. In order to allow a quantum system as the reference, with respect to which the node labels are defined, we need to include it in the description. Recall that $\Hc_{O_1} , \Hc_{O_2}$ are the Hilbert spaces of two quantum reference fields $\hat \chi_1(q), \hat \chi_2(q)$ defined on points in some abstract set $q\in\Mc$.  The Hilbert space is therefore $ \Hc_S \otimes \Hc_{O_1} \otimes \Hc_{O_2}$. The reference used to write the Hamiltonian \eqref{eq:IsingH} is $O_1$.

The classical and QC symmetries are physically interpreted  with respect to commuting reference fields. That is, if we use two reference fields $\hat\chi_1, \hat\chi_2$ to name the nodes, which are such that $[ \hat\chi_i(q), \hat\chi_i(q')    ] = 0$ for all $q,q'$, the Hamiltonian stays invariant under the QP $u$ that relates them, if and only if $u$ is a classical automorphism of the graph or a QC permutation composed from classical automorphisms.

Physically realizing a frame that corresponds to a genuinely quantum graph symmetry, would require that the reference fields used to label the nodes are such that  $[ \hat\chi_i(q), \hat\chi_i(q') ] \neq 0$ for some $q$ and $q'$, so that the $u$ that relates them can be a quantum automorphism of the graph, that is, a BQC permutation. This will generally arise for fields that live in Lorentzian signature spacetimes. Let us now turn to such an example. 

\section{Quantum scalar field on curved spacetime}\label{sec:scalar}

We now study a discretized model for a scalar field $\hphi$ on curved spacetime, and show that it is invariant under quantum permutations that obey a discrete differentiability condition.
The continuum action of a scalar field on a curved spacetime is
\begin{align}
    S = \int \sqrt{-g(x)} \;\text{d}^4x \; \frac12  \partial^\mu\hphi(x) \; g_{\mu\nu}(x) \;\partial^\nu\hphi(x),
\end{align}
where $g_{\mu\nu}$ is the spacetime metric and $g$ its determinant. This action is invariant under diffeomorphisms.

Using similar notation as in Sections \ref{sec:qft} and \ref{sec:localizetwo}, let us define the discrete action from the perspective of $O_1$
\begin{align} 
    S = \sum_{x,x_u,x_v}\Delta\hphi_1(x,x_u)\; \hat g_1(x,x_u,x_v)\; \Delta\hphi_1(x,x_v)
\end{align}
where recall that the subscript $1$ signifies fields as seen by $O_1$. The discrete field derivative is defined as 
\begin{equation}
  \Delta\hphi_1(x,x_u) := \frac{\hphi_1(x_u) - \hphi_1(x)}{\|x_u-x\|}.   
\end{equation}
where $\|\cdot\|$ could be any norm on $\mathbb{R}^n$. The operator $\hat g_1(x,x_u,x_v)$ on $\Hc_S \otimes \Hc_{O_1}$ is a discrete and quantum metric. We take $\hat g_1$ to be a field (a hermitian operator at every point) that is symmetric under the exchange of $x_u$ and $x_v$, as in the classical case. Namely, $g_1(x,x_u,x_v) =g_1(x,x_v,x_u) = g_1(x,x_u,x_v)^\dagger$.

We  assume that $\hat g_1(x,x_u,x_v)$, commutes with $u_{xy}$ for all $x,y$. This assumption is motivated from the examples studied in Section \ref{sec:qft}, Section \ref{sec:localizetwo} and Section \ref{sec:ising}, where we saw that the fields $\hphi_1(x)$ and $\hphi_2(x)$ commute with $u_{xy}$ for all $x,y$.
This can be seen from \eqref{eq:relField} and \eqref{eq:qpfield}, see also see \eqref{eq:QPonscalar}. Intuitively, for every $\hat g_1(x,x_u,x_v)\neq 0$, the points $x_u,x_v$ could be both thought of as ``close" to $x$, or as neighbors of it on some graph or lattice.

We now move to a different coordinate system $\hat\chi_2$. The corresponding QP $u$ is given by \eqref{eq:qpfield}, repeated here for convenience
\begin{align}
    u_{x_2x_1} = \mathbb1_S\otimes\sum_{q} P_1^{qx_1}  \otimes  P_2^{qx_2}.
\end{align}
We assume that $u_{x_2x_1}$ obeys the following \emph{discrete differentiability} condition: given $x,x_u$ such that $\hat g_1(x,x_u,x_v)\neq 0$ at least for one $x_v$, it holds that $u_{xy}u_{x_u y_a} = 0$ unless $y_a$ is of a specific value, denoted by $D_{xy}(x_u)$. This is assumed to be an invertible function, i.e.~$D_{xy}(x_u) \neq D_{xy}(x_u')$ for any $x_u \neq x_u'$. Only some QPs will satisfy this condition, which we call differentiable. 

The transformation rule for the field is given by \eqref{eq:phi12rule}, repeated here for convenience
\begin{align}
    \hphi_1(x)
=\sum_{y} u_{xy}\,\hphi_2(y).
\end{align}
The discrete field derivative transforms as
\begin{align}
\Delta\hphi_1(x,x_u)
&=
\sum_y
u_{xy}\,J_{xy}(x_u)\,
\Delta\hphi_2\!\big(y,D_{xy}(x_u)\big)
\end{align}
where we have defined the (discrete) Jacobian
\begin{align}
J_{xy}(x_u)
:=
\frac{\|D_{xy}(x_u)-y\|}{\|x_u-x\|}.
\end{align}
We prove in Appendix~\ref{ap:scalar} that the action retains its form when moving to the perspective of $O_2$. That is, 
\begin{align}
\label{eq:invAction}
    S = \sum_{y,y_a,y_b}\Delta\hphi_2(y,y_a)\; \hat g_2(y,y_a,y_b)\; \Delta\hphi_2(y,y_b),
\end{align}
with
\begin{align}\label{eq:g2=ug1}
\hat g_2(y,y_a,y_b)
:=
\sum_x u_{xy}\,
J_{xy}\!\big(D^{-1}_{xy}(y_a)\big)\,
J_{xy}\!\big(D^{-1}_{xy}(y_b)\big)
\\\nonumber
\hat g_1\!\big(x,D^{-1}_{xy}(y_a),D^{-1}_{xy}(y_b)\big),
\end{align}
In Appendix~\ref{ap:scalar} we also give (i) a concrete example of such a symmetry of the action, namely a metric and a transformation obeying the differentiability condition (ii) an example of a metric and a transformation that do not obey the differentiability condition, and show that in this case the action does not remain invariant.\footnote{Interestingly, the differentiability condition above is fulfilled by any classical permutation, while this is not the case for quantum permutations.}

Let us summarize. We made a heuristic attempt at a definition for differentiability for quantum permutations: given that $u$ sends $x$ to $y$, $D_{xy}$ dictates that a nearby point $x_u$ will be sent to $y_a = D_{xy}(x_u)$. We saw that the discrete analogue of the quantum field derivative and quantum metric transform analogously to the way tensors transform in general relativity. The invariance of the action then follows. 

Therefore, we have shown that there are QPs that satisfy a discrete sense of differentiability and which correspond to new, genuinely quantum, symmetries of a background independent action.

\section{Discussion}

We have demonstrated that quantum permutations naturally arise in quantum theory as generic quantum reference frame transformations. Importantly, \eqref{eq:microcausality} implies that the non-commutativity of quantum fields at different locations gives rise precisely to the condition that the frame transformation must be a genuinely quantum permutation. 

With this observation as the point of departure, we set out to investigate how genuine QPs non-trivially extend the framework of (ideal) quantum reference frames. We showed that the special subclass of QPs that correspond to superposing classical graph isomorphisms are quantum controlled permutations. Therefore, the genuinely quantum permutations are those that cannot be cast in the form \eqref{eq:qcontrolled}, where the observer's space serves the role of a quantum control space. Genuinely quantum or BQC permutations are interesting because they can encode symmetries present in quantum theory with no classical counterpart: there are classically not-isomorphic graphs that are mapped to each through a BQC permutation (they are quantum isomorphic). Here, we studied quantum permutations in a quantum information language which has allowed us to study the role of this notion of quantum symmetry in a more general context than graphs, e.g~by taking the quantum permutation to act on the eigenbasis elements of observables.

We observed that QC permutations correspond to making a global choice of control basis, the same at each location, while BQC permutations correspond to choosing one control basis per location and so that these bases will not commute on (at least two) different locations (see \eqref{eq:local_basis} and the discussion that follows). In this sense, BQC permutations incorporate the non-commutativity of quantum theory as the non-commutativity of the local choice of quantum reference frame. 

Leveraging this insight, we produced examples of genuinely quantum reference frame transformations. We first demonstrated that QC permutations can be understood as symmetries of the dynamics of a physical system, in particular of any Hamiltonian that is symmetric under (ideal) quantum reference frame transformations studied in the literature. We gave a Hamiltonian that is symmetric under a BQC unitary which `packages together' transformations that control on non-commuting observables on different parts of the real line. We then saw how to generalise the notion of quantum coordinates that appeared in \cite{kabelQuantumCoordinatesLocalisation2025a} to BQC transformations, demonstrating how to localize the state \eqref{eq:toLocalize}, which cannot be localized with QC transformations.

Putting the above together, we saw that the Ising Hamiltonian, when defined on a graph with a genuinely quantum symmetry, is invariant under the corresponding BQC permutation---and that the physical realization of such a frame choice will involve the use of a reference quantum field that does not commute on some of the graph nodes. 

The non-commutativity of field operators is usually related to causality: the principle of microcausality is the fact that quantum fields commute at spacelike separation but in general do not at timelike separation. Given that 
almost all trees (graphs with a partial order) have genuinely quantum symmetries \cite{Junk_2020}, this implies that these new symmetries may become particularly important in a relativistic context of theories defined on spacetimes with Lorentzian signature. Accordingly, we
studied BQC permutations that fulfill a discrete differentiability condition, and showed that, remarkably, they leave invariant a discretization of the action of a scalar field on a curved spacetime, see \eqref{eq:invAction}. This strongly indicates that \emph{ a notion of continuous quantum permutations would provide a definition for `genuinely quantum diffeomorphisms'}; quantum diffeomorphisms that are transformations more general than superpositions of classical diffeomorphisms. Our results set the stage to consider the continuum case, which will be the topic of a future work.

Recently, prototypical invariants under quantum coordinates changes have begun to emerge and their meaning investigated. In  \cite{delaHamette2022quantum}, such an invariant was used to give the first general relativistic definition for indefinite causal ordering (ICO), applicable both when ICO is implemented on an optical bench and when ICO takes place due to a quantum superposition of gravitational fields. General knot invariants in the context of ICO  have since been found \cite{Fedida:2024fna}. Another example is the demonstration that while entanglement and coherence are not generically preserved under quantum (controlled) reference frame transformations, their sum is preserved \cite{cepollaroSumEntanglementSubsystem2025}. The idea of quantum coordinate changes was used in  \cite{Christodoulou:2022knr} to argue that when entanglement generated through the gravitational interaction, the proper distance must be set in superposition (because it can not be made definite with a quantum coordinate change). Most closely related to this work, it has recently been shown that the invariance under quantum controlled permutations rules out parastatistics \cite{mekonnenInvarianceQuantumPermutations2025}  (providing a `reason' for why we see only bosons and fermions). This is an intriguing demonstration of how invariance under quantum reference frames can be used as a guiding principle for physics. 

All of the above works concern a notion of invariance under QC permutations. Furthermore, typically an explicit fixed choice of basis is made (QC transformations for a fixed choice of basis form a group). It is intriguing
 to ask how the picture may change when considering instead invariance under general BQC transformations. Intuitively, it would seem that a \emph{smaller} set of invariants would survive if invariance under genuinely quantum reference frames is posited (instead of invariance under the QC kind). The main challenge here is that the quantum permutations are a sort of quantum group: the group closure of QPs is contained in the group of unitaries of the same dimensionality, whose blocks sum up to one over rows and columns. Whether this closure is the entire group of unitaries with unit sum of rows and columns is unknown, but it is easy to show that it cannot be the entire unitary group. One possibility is to forgo a group structure and use instead a weaker equivalence relation \cite{LupiniMancinskaRoberson2020}, in order to define the quantized version of the `glocal' observables studied in \cite{Broukal:2025cgx}.
 
Finally, we mention another interesting further direction: to investigate whether the theory of magic squares \cite{cuevasQuantumMagicSquares2020} may allow to extend non-ideal QRFs, similarly to how we have done here for ideal QRFs using QPs (magic unitaries). Magic squares are not unitaries but have positive blocks, so the blocks can be written as weighted sums of projections. This seems to point towards some type of coarse-graining of QP blocks.

In closing, we have seen that the theory of quantum permutations significantly extends the notion of symmetry in physics. We have only begun here to uncover the implications.

\section*{Acknowledgments}
Aristotelis Panagiotopoulos is thanked for a mathematician's insight into quantum permutations. MC acknowledges insightful discussions on the role of permutation invariance in physics with Jeremy Butterfield, Eugenio Bianchi and Andrea Di Biagio. OB thanks Emil Broukal and Augustin Vanrietvelde for useful discussions. JB is thanked also for providing comments in an earlier version of this draft. We acknowledge support of the ID\#~62312 `Quantum Information Structure of Spacetime (QISS)'  and ID\#~ 63683  `Without Spacetime (WOST)' grants from the John Templeton Foundation. 

\bibliography{refs}

@article{weberQuantumPermutationMatrices2023,
  title = {Quantum {{Permutation Matrices}}},
  author = {Weber, Moritz},
  year = {2023},
  month = mar,
  journal = {Complex Analysis and Operator Theory},
  volume = {17},
  number = {3},
  pages = {37},
  issn = {1661-8262},
  doi = {10.1007/s11785-023-01335-x},
  urldate = {2023-12-12},
  langid = {english}
}

@misc{delahamettePerspectiveneutralApproachQuantum2021a,
  title = {Perspective-Neutral Approach to Quantum Frame Covariance for General Symmetry Groups},
  author = {{de la Hamette}, Anne-Catherine and Galley, Thomas D. and Hoehn, Philipp A. and Loveridge, Leon and Mueller, Markus P.},
  year = {2021},
  month = oct,
  number = {arXiv:2110.13824},
  eprint = {2110.13824},
  primaryclass = {gr-qc, physics:hep-th, physics:quant-ph},
  publisher = {arXiv},
  doi = {10.48550/arXiv.2110.13824},
  archiveprefix = {arxiv}
}

@misc{Chen_2026,
  title = {Quantum {{Reference Fields Transformations}} in {{Linearized Quantum Gravity}}},
  author = {Chen, Lin-Qing and Giacomini, Flaminia},
  year = 2026,
  month = jun,
  number = {arXiv:2606.09344},
  eprint = {2606.09344},
  primaryclass = {gr-qc},
  publisher = {arXiv},
  doi = {10.48550/arXiv.2606.09344},
  urldate = {2026-07-27},
  archiveprefix = {arXiv}
}

@article{kabelQuantumCoordinatesLocalisation2025a,
  title = {Quantum Coordinates, Localisation of Events, and the Quantum Hole Argument},
  author = {Kabel, Viktoria and {de la Hamette}, Anne-Catherine and Apadula, Luca and Cepollaro, Carlo and Gomes, Henrique and Butterfield, Jeremy and Brukner, {\v C}aslav},
  year = {2025},
  month = apr,
  journal = {Communications Physics},
  volume = {8},
  number = {1},
  pages = {185},
  publisher = {Nature Publishing Group},
  issn = {2399-3650},
  doi = {10.1038/s42005-025-02084-3},
  urldate = {2025-07-04},
  copyright = {2025 The Author(s)},
  langid = {english}
}

@article{krummQuantumReferenceFrame2021a,
  title = {Quantum Reference Frame Transformations as Symmetries and the Paradox of the Third Particle},
  author = {Krumm, Marius and H{\"o}hn, Philipp A. and M{\"u}ller, Markus P.},
  year = {2021},
  month = aug,
  journal = {Quantum},
  volume = {5},
  pages = {530},
  publisher = {Verein zur F{\"o}rderung des Open Access Publizierens in den Quantenwissenschaften},
  doi = {10.22331/q-2021-08-27-530}
}

@article{cepollaroSumEntanglementSubsystem2025,
  title = {Sum of {{Entanglement}} and {{Subsystem Coherence Is Invariant}} under {{Quantum Reference Frame Transformations}}},
  author = {Cepollaro, Carlo and Akil, Ali and Cie{\'s}li{\'n}ski, Pawe{\l} and {de la Hamette}, Anne-Catherine and Brukner, {\v C}aslav},
  year = {2025},
  month = jun,
  journal = {Physical Review Letters},
  volume = {135},
  number = {1},
  pages = {010201},
  publisher = {American Physical Society},
  doi = {10.1103/h6b3-y4vt},
  urldate = {2025-07-04}
}

@article{atseriasQuantumNonsignallingGraph2019,
  title = {Quantum and Non-Signalling Graph Isomorphisms},
  author = {Atserias, Albert and Man{\v c}inska, Laura and Roberson, David E. and {\v S}{\'a}mal, Robert and Severini, Simone and Varvitsiotis, Antonios},
  year = {2019},
  month = may,
  journal = {Journal of Combinatorial Theory, Series B},
  volume = {136},
  pages = {289--328},
  issn = {0095-8956},
  doi = {10.1016/j.jctb.2018.11.002},
  urldate = {2025-04-24}
}

@inproceedings{levandovskyyExistenceQuantumSymmetries2022,
author = {Levandovskyy, Viktor and Eder, Christian and Steenpass, Andreas and Schmidt, Simon and Schanz, Julien and Weber, Moritz},
title = {Existence of Quantum Symmetries for Graphs on Up to Seven Vertices: A Computer based Approach},
year = {2022},
isbn = {9781450386883},
publisher = {Association for Computing Machinery},
address = {New York, NY, USA},
url = {https://doi.org/10.1145/3476446.3535481},
doi = {10.1145/3476446.3535481},
booktitle = {Proceedings of the 2022 International Symposium on Symbolic and Algebraic Computation},
pages = {311–318},
numpages = {8},
location = {Villeneuve-d'Ascq, France},
series = {ISSAC '22}
}

@article{hametteQuantumReferenceFrames2020,
  title = {Quantum Reference Frames for General Symmetry Groups},
  author = {de la Hamette, Anne-Catherine and Galley, Thomas D.},
  year = {2020},
  month = nov,
  journal = {Quantum},
  volume = {4},
  pages = {367},
  publisher = {Verein zur F{\"o}rderung des Open Access Publizierens in den Quantenwissenschaften},
  doi = {10.22331/q-2020-11-30-367},
  urldate = {2025-01-08},
  langid = {british}
}

@article{delahametteQuantumReferenceFrames2023,
  title = {Quantum Reference Frames for an Indefinite Metric},
  author = {{de la Hamette}, Anne-Catherine and Kabel, Viktoria and {Castro-Ruiz}, Esteban and Brukner, {\v C}aslav},
  year = {2023},
  month = aug,
  journal = {Communications Physics},
  volume = {6},
  number = {1},
  pages = {1--15},
  publisher = {Nature Publishing Group},
  issn = {2399-3650},
  doi = {10.1038/s42005-023-01344-4},
  urldate = {2023-09-29},
  copyright = {2023 Springer Nature Limited},
  langid = {english}
}

@article{castro-ruizRelativeSubsystemsQuantum2025,
  title = {Relative Subsystems and Quantum Reference Frame Transformations},
  author = {{Castro-Ruiz}, Esteban and Oreshkov, Ognyan},
  year = {2025},
  month = apr,
  journal = {Communications Physics},
  volume = {8},
  number = {1},
  pages = {187},
  publisher = {Nature Publishing Group},
  issn = {2399-3650},
  doi = {10.1038/s42005-025-02036-x},
  urldate = {2025-07-04},
  copyright = {2025 The Author(s)},
  langid = {english}
}

@article{mekonnenInvarianceQuantumPermutations2025,
  title = {Invariance under Quantum Permutations Rules out Parastatistics},
  author = {Mekonnen, Manuel and Galley, Thomas D. and Mueller, Markus P.},
  year = {2026},
  journal = {Nature Communications},
  volume = {17},
  number = {1},
  pages = {6947},
  eprint = {2502.17576},
  primaryclass = {quant-ph},
  doi = {10.1038/s41467-026-73064-6},
  urldate = {2026-08-13},
  archiveprefix = {arXiv}
}

@misc{zychRelativityQuantumSuperpositions2018,
  title = {Relativity of Quantum Superpositions},
  author = {Zych, Magdalena and Costa, Fabio and Ralph, Timothy C.},
  year = {2018},
  month = sep,
  number = {arXiv:1809.04999},
  eprint = {1809.04999},
  primaryclass = {quant-ph},
  publisher = {arXiv},
  doi = {10.48550/arXiv.1809.04999},
  urldate = {2025-06-24},
  archiveprefix = {arXiv}
}

@article{aharonovQuantumFramesReference1984,
  title = {Quantum Frames of Reference},
  author = {Aharonov, Y. and Kaufherr, T.},
  year = {1984},
  month = jul,
  journal = {Physical Review D},
  volume = {30},
  number = {2},
  pages = {368--385},
  issn = {0556-2821},
  doi = {10.1103/PhysRevD.30.368},
  urldate = {2025-07-03},
  copyright = {http://link.aps.org/licenses/aps-default-license},
  langid = {english}
}

@article{rovelliQuantumReferenceSystems1991,
  title = {Quantum Reference Systems},
  author = {Rovelli, C},
  year = {1991},
  month = feb,
  journal = {Classical and Quantum Gravity},
  volume = {8},
  number = {2},
  pages = {317--331},
  issn = {0264-9381, 1361-6382},
  doi = {10.1088/0264-9381/8/2/012},
  urldate = {2025-07-03},
  langid = {english}
}

@article{cuevasQuantumMagicSquares2020,
  title = {Quantum Magic Squares: Dilations and Their Limitations},
  shorttitle = {Quantum Magic Squares},
  author = {las Cuevas, Gemma De and Drescher, Tom and Netzer, Tim},
  year = {2020},
  month = nov,
  journal = {Journal of Mathematical Physics},
  volume = {61},
  number = {11},
  eprint = {1912.07332},
  primaryclass = {quant-ph},
  pages = {111704},
  issn = {0022-2488, 1089-7658},
  doi = {10.1063/5.0022344},
  urldate = {2025-06-16},
  archiveprefix = {arXiv}
}

@article{hohnEquivalenceApproachesRelational2021,
  title = {Equivalence of {{Approaches}} to {{Relational Quantum Dynamics}} in {{Relativistic Settings}}},
  author = {H{\"o}hn, Philipp A. and Smith, Alexander R. H. and Lock, Maximilian P. E.},
  year = {2021},
  month = jun,
  journal = {Frontiers in Physics},
  volume = {9},
  publisher = {Frontiers},
  issn = {2296-424X},
  doi = {10.3389/fphy.2021.587083},
  urldate = {2025-07-04},
  langid = {english}
}

@article{hohnTrinityRelationalQuantum2021,
  title = {Trinity of Relational Quantum Dynamics},
  author = {H{\"o}hn, Philipp A. and Smith, Alexander R. H. and Lock, Maximilian P. E.},
  year = {2021},
  month = sep,
  journal = {Physical Review D},
  volume = {104},
  number = {6},
  pages = {066001},
  publisher = {American Physical Society},
  doi = {10.1103/PhysRevD.104.066001},
  urldate = {2025-07-04}
}

@article{woronowiczCompactMatrixPseudogroups1987,
  title = {Compact Matrix Pseudogroups},
  author = {Woronowicz, S. L.},
  year = 1987,
  month = dec,
  journal = {Communications in Mathematical Physics},
  volume = {111},
  number = {4},
  pages = {613--665},
  issn = {0010-3616, 1432-0916},
  doi = {10.1007/BF01219077},
  urldate = {2025-11-04},
  copyright = {http://www.springer.com/tdm},
  langid = {english}
}

@article{Oreshkov:2011er,
    author = "Oreshkov, Ognyan and Costa, Fabio and Brukner, Caslav",
    title = "{Quantum correlations with no causal order}",
    eprint = "1105.4464",
    archivePrefix = "arXiv",
    primaryClass = "quant-ph",
    doi = "10.1038/ncomms2076",
    journal = "Nature Commun.",
    volume = "3",
    pages = "1092",
    year = "2012"
}

@article{PhysRevX.8.011047,
  title = {Dynamics of Quantum Causal Structures},
  author = {Castro-Ruiz, Esteban and Giacomini, Flaminia and Brukner, \ifmmode \check{C}\else \v{C}\fi{}aslav},
  journal = {Phys. Rev. X},
  volume = {8},
  issue = {1},
  pages = {011047},
  numpages = {15},
  year = {2018},
  month = {Mar},
  publisher = {American Physical Society},
  doi = {10.1103/PhysRevX.8.011047},
  url = {https://link.aps.org/doi/10.1103/PhysRevX.8.011047}
}

@article{delaHamette2022quantum,
  title = {Indefinite Causal Order and Quantum Coordinates},
  author = {de la Hamette, Anne-Catherine and Kabel, Viktoria and Christodoulou, Marios and Brukner, \ifmmode \check{C}\else \v{C}\fi{}aslav},
  journal = {Phys. Rev. Lett.},
  volume = {135},
  issue = {14},
  pages = {141402},
  numpages = {8},
  year = {2025},
  month = {Oct},
  publisher = {American Physical Society},
  doi = {10.1103/bnkn-4p3f},
  url = {https://link.aps.org/doi/10.1103/bnkn-4p3f}
}

@article{giacominiQRF2019,
	author = {Giacomini, Flaminia and Castro-Ruiz, Esteban and Brukner, {\v C}aslav},
	date = {2019/01/30},
	doi = {10.1038/s41467-018-08155-0},
	eprint = {1712.07207},
	archiveprefix = {arXiv},
	primaryclass = {quant-ph},
	id = {Giacomini2019},
	isbn = {2041-1723},
	journal = {Nature Communications},
	number = {1},
	pages = {494},
	title = {Quantum mechanics and the covariance of physical laws in quantum reference frames},
	url = {https://doi.org/10.1038/s41467-018-08155-0},
	volume = {10},
	year = {2019}}

@article{Zych:2015fka,
    author = "Zych, Magdalena and Brukner, {\v{C}}aslav",
    title = "{Quantum formulation of the Einstein Equivalence Principle}",
    eprint = "1502.00971",
    archivePrefix = "arXiv",
    primaryClass = "gr-qc",
    doi = "10.1038/s41567-018-0197-6",
    journal = "Nature Phys.",
    volume = "14",
    number = "10",
    pages = "1027--1031",
    year = "2018"
}

@InProceedings{Hardy:2019cef,
author="Hardy, Lucien",
editor="Finster, Felix
and Giulini, Domenico
and Kleiner, Johannes
and Tolksdorf, J{\"u}rgen",
title="Implementation of the Quantum Equivalence Principle",
booktitle="Progress and Visions in Quantum Theory in View of Gravity",
year="2020",
publisher="Springer International Publishing",
address="Cham",
pages="189--220",
isbn="978-3-030-38941-3",
doi={10.1007/978-3-030-38941-3_8}
}

@article{Fedida:2024fna,
    author = "Fedida, Samuel and de la Hamette, Anne-Catherine and Kabel, Viktoria and Brukner, {\v{C}}aslav",
    title = "{Knot invariants and indefinite causal order}",
    eprint = "2409.11448",
    archivePrefix = "arXiv",
    primaryClass = "gr-qc",
    doi = "10.22331/q-2025-10-06-1875",
    journal = "Quantum",
    volume = "9",
    pages = "1875",
    year = "2025"
}

@article{Christodoulou:2022knr,
    author = "Christodoulou, Marios and Di Biagio, Andrea and Howl, Richard and Rovelli, Carlo",
    title = "{Gravity entanglement, quantum reference systems, degrees of freedom}",
    eprint = "2207.03138",
    archivePrefix = "arXiv",
    primaryClass = "quant-ph",
    doi = "10.1088/1361-6382/acb0aa",
    journal = "Class. Quant. Grav.",
    volume = "40",
    number = "4",
    pages = "047001",
    year = "2023"
}

@article{Chiribella:2009lvz,
    author = "Chiribella, Giulio and D'Ariano, Giacomo Mauro and Perinotti, Paolo and Valiron, Benoit",
    title = "{Quantum computations without definite causal structure}",
    eprint = "0912.0195",
    archivePrefix = "arXiv",
    primaryClass = "quant-ph",
    doi = "10.1103/PhysRevA.88.022318",
    journal = "Phys. Rev. A",
    volume = "88",
    number = "2",
    pages = "022318",
    year = "2013"
}

@article{Rovelli:1995fv,
    author = "Rovelli, Carlo",
    title = "{Relational quantum mechanics}",
    eprint = "quant-ph/9609002",
    archivePrefix = "arXiv",
    doi = "10.1007/BF02302261",
    journal = "Int. J. Theor. Phys.",
    volume = "35",
    pages = "1637--1678",
    year = "1996"
}

@article{Junk_2020,
  title = {Almost All Trees Have Quantum Symmetry},
  author = {Junk, Luca and Schmidt, Simon and Weber, Moritz},
  year = 2020,
  month = oct,
  journal = {Archiv der Mathematik},
  volume = {115},
  number = {4},
  pages = {367--378},
  issn = {1420-8938},
  doi = {10.1007/s00013-020-01476-x}
}

@article{ErdosRenyi1963,
  author  = {P. Erd\H{o}s and A. R{\'e}nyi},
  title   = {Asymmetric Graphs},
  journal = {Acta Mathematica Academiae Scientiarum Hungaricae},
  volume  = {14},
  number  = {3--4},
  pages   = {295--315},
  year    = {1963},
  doi     = {10.1007/BF01895716}
}

@article{LupiniMancinskaRoberson2020,
  author  = {Martino Lupini and Laura Man{\v{c}}inska and David E. Roberson},
  title   = {Nonlocal Games and Quantum Permutation Groups},
  journal = {Journal of Functional Analysis},
  volume  = {279},
  number  = {5},
  pages   = {108592},
  year    = {2020},
  doi     = {10.1016/j.jfa.2020.108592}
}

@article{Jung_2020,
  title = {Models of Quantum Permutations},
  author = {Jung, Stefan and Weber, Moritz},
  year = 2020,
  month = aug,
  journal = {Journal of Functional Analysis},
  volume = {279},
  number = {2},
  pages = {108516},
  issn = {0022-1236},
  doi = {10.1016/j.jfa.2020.108516},
  urldate = {2025-11-04}
}

@article{Castro-Ruiz_2020,
  title = {Quantum Clocks and the Temporal Localisability of Events in the Presence of Gravitating Quantum Systems},
  author = {{Castro-Ruiz}, Esteban and Giacomini, Flaminia and Belenchia, Alessio and Brukner, {\v C}aslav},
  year = 2020,
  month = may,
  journal = {Nature Communications},
  volume = {11},
  number = {1},
  pages = {2672},
  publisher = {Nature Publishing Group},
  issn = {2041-1723},
  doi = {10.1038/s41467-020-16013-1},
  urldate = {2022-09-13},
  copyright = {2020 The Author(s)},
  langid = {english}
}

@article{AliAhmad_2022,
  title = {Quantum {{Relativity}} of {{Subsystems}}},
  author = {Ali Ahmad, Shadi and Galley, Thomas D. and H{\"o}hn, Philipp A. and Lock, Maximilian P. E. and Smith, Alexander R. H.},
  year = 2022,
  month = apr,
  journal = {Physical Review Letters},
  volume = {128},
  number = {17},
  pages = {170401},
  publisher = {American Physical Society},
  doi = {10.1103/PhysRevLett.128.170401},
  urldate = {2026-04-01}
}

@article{Loveridge_2018,
  title = {Symmetry, {{Reference Frames}}, and {{Relational Quantities}} in {{Quantum Mechanics}}},
  author = {Loveridge, Leon and Miyadera, Takayuki and Busch, Paul},
  year = 2018,
  month = feb,
  journal = {Foundations of Physics},
  volume = {48},
  number = {2},
  pages = {135--198},
  issn = {1572-9516},
  doi = {10.1007/s10701-018-0138-3},
  urldate = {2026-04-01},
  langid = {english}
}

@article{Vanrietvelde2020changeof,
  title = {A change of perspective: switching quantum reference frames via a perspective-neutral framework},
  author = {Vanrietvelde, Augustin and Hoehn, Philipp A. and Giacomini, Flaminia and Castro-Ruiz, Esteban},
  journal = {Quantum},
  volume = {4},
  pages = {225},
  year = {2020},
  month = jan,
  issn = {2521-327X},
  doi = {10.22331/q-2020-01-27-225},
  url = {https://doi.org/10.22331/q-2020-01-27-225}
}

@misc{Broukal:2025cgx,
    author = "Broukal, Emil and Di Biagio, Andrea and Bianchi, Eugenio and Christodoulou, Marios",
    title = "{Observables are glocal}",
    eprint = "2508.02346",
    archivePrefix = "arXiv",
    primaryClass = "gr-qc",
    month = "8",
    year = "2025"
}

@article{giacominiQuantumSuperpositionSpacetimes2022,
  title = {Quantum Superposition of Spacetimes Obeys Einstein's Equivalence Principle},
  author = {Giacomini, Flaminia and Brukner, {\v C}aslav},
  year = {2022},
  journal = {AVS Quantum Science},
  volume = {4},
  pages = {015601},
  doi = {10.1116/5.0070018},
  eprint = {2109.01405},
  archivePrefix = {arXiv},
  primaryClass = {gr-qc}
}

@misc{Hoehn_2023,
  title = {Matter Relative to Quantum Hypersurfaces},
  author = {Hoehn, Philipp A. and Russo, Andrea and Smith, Alexander R. H.},
  year = 2023,
  month = nov,
  number = {arXiv:2308.12912},
  eprint = {2308.12912},
  primaryclass = {quant-ph},
  publisher = {arXiv},
  doi = {10.48550/arXiv.2308.12912},
  urldate = {2026-07-27},
  archiveprefix = {arXiv}
}

\appendix
\section{Proof of the appearance of quantum permutations in the quantum mechanical setting}\label{ap:proof1stq}

From 1, $\bra {x_1}\hat X_2 \ket{x_1'} = \delta_{x_1x_1'}\,\hat X_{2,x_1}$, so $\hat X_2 = \sum_{x_1} \ket{x_1}\bra{x_1}\otimes \hat X_{2,x_1}$.

Any $\hat X_{2,x_1}$ is hermitian, thus can be written as
$\hat X_{2,x_1} = \sum_{x_2} x_2\; u_{x_1x_2}$,
where the operators
\(u_{x_1x_2} = (u_{x_1x_2})^\dagger = (u_{x_1x_2})^2\)
are orthogonal projections that satisfy
\(\sum_{x_2} u_{x_1x_2} = \mathbb1_{O_2}\).

From 2 it follows that the $x_2$ eigenspace of $\hat X_{2,x_1}$ and the $x_2$ eigenspace of $\hat X_{2,x_1'}$ for $x_1\neq x_1'$ are orthogonal. Therefore
$u_{x_1x_2}u_{x_1'x_2} = \delta_{x_1x_1'}\,u_{x_1x_2}$.
Denote now $u_{x_2} := \sum_{x_1} u_{x_1x_2}$ and $\text{r}(\cdot)$ the rank of a projection.
From the orthogonality we have just obtained,
$\text{r}(u_{x_2}) = \sum_{x_1} \text{r}(u_{x_1x_2})$.
Summing over $x_2$ one gets
$\sum_{x_2} \text{r}(u_{x_2}) = \sum_{x_1} \sum_{x_2} \text{r}(u_{x_1x_2})$.
Since $u_{x_1x_2}$ are orthogonal projections and
\(\sum_{x_2} u_{x_1x_2} = \mathbb1_{O_2}\),
it follows that $\sum_{x_2} \text{r}(u_{x_1x_2})=d$ for all $x_1$.
Thus $\sum_{x_2} \text{r}(u_{x_2}) = \sum_{x_1} d = Nd$.
Since the rank of a projection can is maximally the dimension of the space, we have here $N$ numbers $\leq d$ that sum up to $Nd$.
This means that $\text{r}(u_{x_2}) = d$ for all $x_2$.
As the only maximal rank projection is the identity,
$u_{x_2} = \sum_{x_1} u_{x_1x_2} = \mathbb1_{O_2}$ for all $x_2$.

Now returning to the relative value observable,
\begin{align}
\hat X_2
&= \sum_{x_1} \ket{x_1}\bra{x_1}\otimes \hat X_{2,x_1} \\\nonumber
&= \sum_{x_1} \ket{x_1}\bra{x_1}\otimes \sum_{x_2} x_2\,u_{x_1x_2} \\\nonumber
&=\sum_{x_1,x_1',x_2} x_2\; \ket{x_1}\bra{x_1'}_S \otimes u_{x_1x_2}u_{x_1'x_2} \\\nonumber
&=u \; \hat X_1 \; u^\dagger,
\end{align}
with
\begin{align}
u = \sum_{x_1,x_2} \ket{x_1}\bra{x_2}_S \otimes u_{x_1x_2}.
\end{align}

The opposite direction follows straightforwardly.

\section{Localizing two particles simultaneously}\label{ap:localizetwo}

The non-relativistic limit lets us write the annihilation operators approximately as
$a_1(x)=\frac{1}{\sqrt{2}}\left(\frac{\sqrt{m c^{2}}}{\hbar}\,\hphi_{A,1}(x)+\frac{i}{\sqrt{m c^{2}}}\,\hat\Pi_{A,1}(x)\right)$ with the conjugate momentum of $A$ relative to $O_1$ denoted $\hat\Pi_{A,1}(x)$, and same for $b$ and $B$. The reason is that in this regime, the Klein--Gordon equation that \(A,B\) satisfy,
\(
\left(\frac{1}{c^{2}}\partial_{t}^{2}-\nabla^{2}+\left(\frac{mc}{\hbar}\right)^{2}\right)\hphi_1(x)=0,
\)
looks like a family of equations for independent simple harmonic oscillators at every \(x\) because the mass term \(\left(\frac{mc}{\hbar}\right)^{2}\hphi_1(x)\) dominates the gradient term \(\nabla^{2}\hphi_1(x)\). Here, since we work in the discrete, \(\partial_t\) and \(\nabla\) are understood as their standard finite-difference analogues.

The same transformation rule \eqref{eq:phi12rule} that we found for a field follows identically for its conjugate momentum. It is thus straightforward to see that it also holds for $a$ and $b$.

The state can be written as
\begin{align}
\ket\psi =
    &\frac{1}{\sqrt2}\,a_1^\dagger(x_A)\,
    \frac{b_1^\dagger(x_B)+b_1^\dagger(x_B')}{\sqrt2}\;
    \ket\Omega_A\ket\Omega_B\ket{\phi}_{O_1O_2}
    \nonumber\\
    &+\frac{1}{\sqrt2}\,a_1^\dagger(x_A')\,
    \frac{b_1^\dagger(x_B)-b_1^\dagger(x_B')}{\sqrt2}\;
    \ket\Omega_A\ket\Omega_B\ket{\phi^\perp}_{O_1O_2}.
\end{align}

From the definiteness of \(O_1\), it holds that
\(a_1(y_A)=a(q_A)\otimes\mathbb{1}_{BO_1O_2}\) and
\(b_1(y_B)=b(q_B)\otimes\mathbb{1}_{AO_1O_2}\).
Hence \(a_1^\dagger(\cdot)\) commutes with \(b_1^\dagger(\cdot)\), and we may write
\begin{align}\label{eq:b1a1}
\ket\psi =
    &\frac{b_1^\dagger(x_B)+b_1^\dagger(x_B')}{2}\;
    a_1^\dagger(x_A)\,\ket\Omega_A\ket\Omega_B\ket{\phi}_{O_1O_2}
    \nonumber\\
    &+\frac{b_1^\dagger(x_B)-b_1^\dagger(x_B')}{2}\;
    a_1^\dagger(x_A')\,\ket\Omega_A\ket\Omega_B\ket{\phi^\perp}_{O_1O_2}.
\end{align}

Applying the transformation rule for \(A\),
\begin{align}
a_1^\dagger(x_A)
&=a_2^\dagger(x_A)\ket{\phi}\bra{\phi}
 +a_2^\dagger(x_A')\ket{\phi^\perp}\bra{\phi^\perp},
\nonumber\\
a_1^\dagger(x_A')
&=a_2^\dagger(x_A)\ket{\phi^\perp}\bra{\phi^\perp}
 +a_2^\dagger(x_A')\ket{\phi}\bra{\phi},
\end{align}
So the state is
\begin{align}
\ket\psi =
    &\frac{b_1^\dagger(x_B)+b_1^\dagger(x_B')}{2}\;
    a_2^\dagger(x_A)\,\ket\Omega_A\ket\Omega_B\ket{\phi}_{O_1O_2}
    \nonumber\\
    &+\frac{b_1^\dagger(x_B)-b_1^\dagger(x_B')}{2}\;
    a_2^\dagger(x_A)\,\ket\Omega_A\ket\Omega_B\ket{\phi^\perp}_{O_1O_2}.
\end{align}
For the same reason above, \(b_1^\dagger(\cdot)\) commutes with \(a_2^\dagger(\cdot)\). Thus we may write
\begin{align}
\ket\psi =
    &\frac12\,a_2^\dagger(x_A)\,
    \big(b_1^\dagger(x_B)+b_1^\dagger(x_B')\big)\,
    \ket\Omega_A\ket\Omega_B\ket{\phi}_{O_1O_2}
    \nonumber\\
    &+\frac12\,a_2^\dagger(x_A)\,
    \big(b_1^\dagger(x_B)-b_1^\dagger(x_B')\big)\,
    \ket\Omega_A\ket\Omega_B\ket{\phi^\perp}_{O_1O_2}.
\end{align}

Applying the transformation rule for \(B\),
\begin{align}
b_1^\dagger(x_B)
&=b_2^\dagger(x_B)\ket{+}\bra{+}
 +b_2^\dagger(x_B')\ket{-}\bra{-},
\nonumber\\
b_1^\dagger(x_B')
&=b_2^\dagger(x_B)\ket{-}\bra{-}
 +b_2^\dagger(x_B')\ket{+}\bra{+},
\end{align}
so we have
\begin{align}
b_1^\dagger(x_B)+b_1^\dagger(x_B')
&=\big(b_2^\dagger(x_B)+b_2^\dagger(x_B')\big)
\big(\ket{+}\bra{+}+\ket{-}\bra{-}\big)
\nonumber\\
&=\big(b_2^\dagger(x_B)+b_2^\dagger(x_B')\big),
\\
b_1^\dagger(x_B)-b_1^\dagger(x_B')
&=\big(b_2^\dagger(x_B)-b_2^\dagger(x_B')\big)
\big(\ket{+}\bra{+}-\ket{-}\bra{-}\big),
\\
&=\big(b_2^\dagger(x_B)-b_2^\dagger(x_B')\big)\big(\ket{\phi}\bra{\phi^\perp}+\ket{\phi^\perp}\bra{\phi}\big).
\label{eq:pm_flip_app}
\end{align}
Hence
\begin{align}
\ket\psi =
    &\frac12\,a_2^\dagger(x_A)\,
    \big(b_2^\dagger(x_B)+b_2^\dagger(x_B')\big)\,
    \ket\Omega_A\ket\Omega_B\ket{\phi}_{O_1O_2}
    \nonumber\\
    &+\frac12\,a_2^\dagger(x_A)\,
    \big(b_2^\dagger(x_B)-b_2^\dagger(x_B')\big)\,
    \ket\Omega_A\ket\Omega_B\ket{\phi}_{O_1O_2}.
    \\\nonumber
=&a_2^\dagger(x_A)\,b_2^\dagger(x_B)\,
\ket\Omega_A\ket\Omega_B\ket{\phi}_{O_1O_2}.
\end{align}

Let us now show that no $QC$ permutation can do this. Assume, for contradiction, that a QC transformation from \(O_1\) to \(O_2\)
can map the state to a state in which both particles are created at definite spacetime points. In a QC permutation/diffeomorphism the same control decomposition is used for all particles. Thus there exist orthogonal projectors \(\{\pi_k\}\) on \(\Hc_{O_1}\otimes\Hc_{O_2}\) and bijections \(f_k:\Mc\to\mathbb{R}^4\) such that
\begin{align}
a_1^\dagger(x)
&=\sum_k a_2^\dagger\!\big(f_k(x)\big)\,\pi_k,
&
b_1^\dagger(x)
&=\sum_k b_2^\dagger\!\big(f_k(x)\big)\,\pi_k.
\label{eq:qc_rule}
\end{align}
Applying \eqref{eq:qc_rule} to \(A\) in \eqref{eq:b1a1} and bringing \(a_2^\dagger\) to the left gives
\begin{align}
&\ket\psi =\\\nonumber
&\sum_k \frac12\,
a_2^\dagger\!\big(f_k(x_A)\big)\,
\big(b_1^\dagger(x_B)+b_1^\dagger(x_B')\big)
\ket{\Omega_A}\ket{\Omega_B}\pi_k\ket{\phi}+
\nonumber\\
&
\sum_k \frac12\,
a_2^\dagger\!\big(f_k(x_A')\big)\,
\big(b_1^\dagger(x_B)-b_1^\dagger(x_B')\big)
\ket{\Omega_A}\ket{\Omega_B}\pi_k\ket{\phi^\perp}.
\end{align}

Applying \eqref{eq:qc_rule} to \(B\),
\begin{align}
\ket\psi
=\sum_k \Bigg[&
\frac12\,
a_2^\dagger\!\big(f_k(x_A)\big)\,
\big(b_2^\dagger(f_k(x_B))+b_2^\dagger(f_k(x_B'))\big)\,
\nonumber\\
&\hspace{2.5em}\ket{\Omega_A}\ket{\Omega_B}\;\pi_k\ket{\phi}
\nonumber\\
+
&\frac12\,
a_2^\dagger\!\big(f_k(x_A')\big)\,
\big(b_2^\dagger(f_k(x_B))-b_2^\dagger(f_k(x_B'))\big)\,
\nonumber\\
&\hspace{2.5em}\ket{\Omega_A}\ket{\Omega_B}\;\pi_k\ket{\phi^\perp}
\Bigg].
\end{align}
If the final state had both particles definite (e.g.\ proportional to
\(a_2^\dagger(\tilde x_A)b_2^\dagger(\tilde x_B)\ket{\Omega_A}\ket{\Omega_B}\ket{\tilde\phi}\)), then for every \(k\) with nonzero amplitude we would need
\begin{align}
f_k(x_A)=f_k(x_A')=\tilde x_A,
\qquad
f_k(x_B)=f_k(x_B')=\tilde x_B.
\label{eq:qc_needed}
\end{align}
But each \(f_k\) is a permutation, so it cannot satisfy this. Therefore no QC permutation can make both particles simultaneously definite for the state $\ket\psi$.

\section{Scalar field toy model}\label{ap:scalar}
In this Appendix we (1) prove the invariance of the action under quantum permutations obeying a differentiability condition (2) present an example of a metric and a transformation that obeying the differentiability condition (3) present a counter-example of a metric and a transformation that do not obey the differentiability condition, and show that in this case the action does not remain invariant. 

\subsection{Proof of action invariance}

Let us prove something more general than what was stated in the text, namely that the action stays invariant also when the finite difference is divided by the coordinate distance:
\begin{align}
\Delta\hphi_1(x,x_u)
:=
\frac{\hphi_1(x_u)-\hphi_1(x)}{\|x_u-x\|},
\end{align}
with \(\|\cdot\|\) any norm on \(\mathbb R^n\). The main text focuses on the case where the norms are absent, or equivalently are all unit. From the differentiability condition, for every pair \((x,y)\) and \(x_u\) with $\hat g_1(x,x_u,x_v)\neq 0$ at least for one $x_v$, we have a definite image \(D_{xy}(x_u)\). We thus define
\begin{align}
J_{xy}(x_u)
:=
\frac{\|D_{xy}(x_u)-y\|}{\|x_u-x\|}.
\end{align}

Begin by proving the transformation rule for the finite differences:
\begin{align}\label{eq:begin_finite_difference}
\Delta\hphi_1(x,x_u)
&=
\frac{\hphi_1(x_u)-\hphi_1(x)}{\|x_u-x\|}
\\\nonumber
&=
\sum_{y_a}
\frac{\big(u_{x_u y_a}-u_{xy_a}\big)\hphi_2(y_a)}{\|x_u-x\|}.
\end{align}
Multiplying on the left by \(u_{xy}\), and using
\(u_{xy}u_{x_u y_a}=0\) unless \(y_a=D_{xy}(x_u)\), together with
\(u_{xy}u_{xy_a}=0\) unless \(y_a=y\), gives
\begin{align}
u_{xy}\,\Delta\hphi_1(x,x_u)
&=
\frac{
u_{xy}u_{x_u\,D_{xy}(x_u)}\hphi_2\!\big(D_{xy}(x_u)\big)
-u_{xy}\hphi_2(y)
}{
\|x_u-x\|
}.
\end{align}
Using
\begin{align}
u_{xy}u_{x_u\,D_{xy}(x_u)}
=\sum_{y'}u_{xy}u_{x_u y'}
=u_{xy},
\end{align}
we obtain
\begin{align}
u_{xy}\,\Delta\hphi_1(x,x_u)
&=
u_{xy}\,
\frac{\hphi_2(D_{xy}(x_u))-\hphi_2(y)}{\|x_u-x\|}
\\\nonumber
&=
u_{xy}\,
\frac{\|D_{xy}(x_u)-y\|}{\|x_u-x\|}
\Delta\hphi_2\!\big(y,D_{xy}(x_u)\big)
\\\nonumber
&=
u_{xy}\,J_{xy}(x_u)\,
\Delta\hphi_2\!\big(y,D_{xy}(x_u)\big).
\end{align}
Summing over \(y\),
\begin{align}\label{eq:app-delta-pushed-normed}
\Delta\hphi_1(x,x_u)
&=
\sum_y
u_{xy}\,J_{xy}(x_u)\,
\Delta\hphi_2\!\big(y,D_{xy}(x_u)\big)
\\\nonumber
&=
\sum_y
J_{xy}(x_u)\,
\Delta\hphi_2\!\big(y,D_{xy}(x_u)\big)\,
u_{xy}.
\end{align}
The second equality follows from repeating the calculation with the QP entries to the right of the field operators.

Now start from
\begin{align}
S
=\sum_{x,x_u,x_v}&
\Delta\hphi_1(x,x_u)\;
\hat g_1(x,x_u,x_v)\;
\Delta\hphi_1(x,x_v).
\end{align}
Substituting \eqref{eq:app-delta-pushed-normed},
\begin{align}
S
=\sum_{\substack{x,x_u,x_v \\ y,y'}}
&
J_{xy}(x_u)\,
\Delta\hphi_2\!\big(y,D_{xy}(x_u)\big)
\\\nonumber
&u_{xy}\,
\hat g_1(x,x_u,x_v)\,
u_{xy'}
\\\nonumber
&J_{xy'}(x_v)\,
\Delta\hphi_2\!\big(y',D_{xy'}(x_v)\big).
\end{align}
By assumption the operators \(u_{xy},u_{xy'}\) commute with \(\hat g_1(x,x_u,x_v)\). Hence for \(y\neq y'\),
\begin{align}
u_{xy}\,
\hat g_1(x,x_u,x_v)\,
u_{xy'}
=
u_{xy}u_{xy'}\,
\hat g_1(x,x_u,x_v)
=
0,
\end{align}
since \(u_{xy}u_{xy'}=0\) for \(y\neq y'\). Thus,
\begin{align}
S
=\sum_{x,x_u,x_v,y}
&
J_{xy}(x_u)\,
\Delta\hphi_2\!\big(y,D_{xy}(x_u)\big)
\\\nonumber
&u_{xy}\,
\hat g_1(x,x_u,x_v)
\\\nonumber
&J_{xy}(x_v)\,
\Delta\hphi_2\!\big(y,D_{xy}(x_v)\big)
\\\nonumber
=\sum_{x,y}\sum_{y_a,y_b}
&
J_{xy}\!\big(D^{-1}_{xy}(y_a)\big)\,
\Delta\hphi_2(y,y_a)
\\\nonumber
&u_{xy}\,
\hat g_1\!\big(x,D^{-1}_{xy}(y_a),D^{-1}_{xy}(y_b)\big)
\\\nonumber
&J_{xy}\!\big(D^{-1}_{xy}(y_b)\big)\,
\Delta\hphi_2(y,y_b),
\end{align}
where in the second equality we changed variables from \(x_u,x_v\) to
\begin{align}
    y_a=D_{xy}(x_u),
    \qquad
    y_b=D_{xy}(x_v),
\end{align}
using the invertibility of \(D_{xy}\).

We may therefore define
\begin{align}\label{eq:g2-general-derivative}
\hat g_2(y,y_a,y_b)
:=
\sum_x u_{xy}\,
J_{xy}\!\big(D^{-1}_{xy}(y_a)\big)\,
J_{xy}\!\big(D^{-1}_{xy}(y_b)\big)
\\\nonumber
\hat g_1\!\big(x,D^{-1}_{xy}(y_a),D^{-1}_{xy}(y_b)\big),
\end{align}
so that
\begin{align}
S
=
\sum_{y,y_a,y_b}
\Delta\hphi_2(y,y_a)\;
\hat g_2(y,y_a,y_b)\;
\Delta\hphi_2(y,y_b).
\end{align}
Thus the action retains its form, with the derivative factors \(J_{xy}(x_u)\) entering the transformation laws in an analogous manner to GR.

Moreover, since
\begin{align}
\hat g_1(x,x_u,x_v)
=
\hat g_1(x,x_v,x_u)
=
\hat g_1(x,x_u,x_v)^\dagger,
\end{align}
the same holds for \(\hat g_2\). Indeed,
\begin{align}
\hat g_2(y,y_a,y_b)^\dagger
&=
\sum_x u_{xy}\,
J_{xy}\!\big(D^{-1}_{xy}(y_a)\big)\,
J_{xy}\!\big(D^{-1}_{xy}(y_b)\big)
\\\nonumber
&\qquad\qquad
\hat g_1\!\big(x,D^{-1}_{xy}(y_a),D^{-1}_{xy}(y_b)\big)^\dagger
\\\nonumber
&=
\sum_x u_{xy}\,
J_{xy}\!\big(D^{-1}_{xy}(y_a)\big)\,
J_{xy}\!\big(D^{-1}_{xy}(y_b)\big)
\\\nonumber
&\qquad\qquad
\hat g_1\!\big(x,D^{-1}_{xy}(y_a),D^{-1}_{xy}(y_b)\big)
\\\nonumber
&=
\hat g_2(y,y_a,y_b),
\end{align}
and since \(\hat g_1\) is symmetric in its last two arguments, so is \(\hat g_2\). Hence
\begin{align}
    \hat g_2(y,y_a,y_b)
    =
    \hat g_2(y,y_a,y_b)^\dagger
    =
    \hat g_2(y,y_b,y_a).
\end{align}
Commutation of $\hat g_2(y,y_a,y_b)$ with $u_{xy}$ for all $x$ also follows immediately from the definition.

\subsection{Concrete example}

A concrete example for the differentiability condition is obtained from four disjoint isomorphic copies of the same graph \(\Lambda\) comprising together a graph $\Gamma$. Denote the corresponding vertices by \(1_i,2_i,3_i,4_i\), where \(i\) runs over the vertices of \(\Lambda\), and take any \(\hat g_1(x,x_u,x_v)\) that is zero unless \(x\sim_\Gamma x_u,x_v\). Define the following BQC:
\begin{align}
    &u_{1_i\,1_i}=\pi,\qquad u_{1_i\,2_i}=\mathbb 1-\pi,
    \\\nonumber
    &u_{2_i\,1_i}=\mathbb 1-\pi,\qquad u_{2_i\,2_i}=\pi,
    \\\nonumber
    &u_{3_i\,3_i}=\pi',\qquad u_{3_i\,4_i}=\mathbb 1-\pi',
    \\\nonumber
    &u_{4_i\,3_i}=\mathbb 1-\pi',\qquad u_{4_i\,4_i}=\pi',
\end{align}
for every \(i\), with all other entries equal to zero, where \(\pi,\pi'\) are two non-commuting orthogonal projections on \(\Hc_{O_1}\otimes\Hc_{O_2}\).

This \(u\) is differentiable. Indeed, if \(1_j\sim_\Gamma 1_i\), then
\begin{align}
    D_{1_i\,1_i}(1_j)=1_j,
    \qquad
    D_{1_i\,2_i}(1_j)=2_j,
\end{align}
and similarly, if \(2_j\sim_\Gamma 2_i\),
\begin{align}
    D_{2_i\,2_i}(2_j)=2_j,
    \qquad
    D_{2_i\,1_i}(2_j)=1_j.
\end{align}
Likewise, for \(3_j\sim_\Gamma 3_i\) and \(4_j\sim_\Gamma 4_i\),
\begin{align}
    D_{3_i\,3_i}(3_j)=3_j,
    \qquad
    D_{3_i\,4_i}(3_j)=4_j,
    \\\nonumber
    D_{4_i\,4_i}(4_j)=4_j,
    \qquad
    D_{4_i\,3_i}(4_j)=3_j.
\end{align}

\subsection{Counter example: a square}

An example of a QP that is not differentiable and does not preserve the action above is the following QP on 4 elements:
\begin{align}
    &u_{11}=\pi,\qquad u_{12}=\mathbb 1-\pi,
    \\\nonumber
    &u_{21}=\mathbb 1-\pi,\qquad u_{22}=\pi,
    \\\nonumber
    &u_{33}=\pi',\qquad u_{34}=\mathbb 1-\pi',
    \\\nonumber
    &u_{43}=\mathbb 1-\pi',\qquad u_{44}=\pi',
\end{align}
with all other entries equal to zero, where again \(\hat g_1(x,x_u,x_v)=0\) unless \(x\sim_\Gamma x_u,x_v\), with \(\Gamma\) the square graph.

Indeed, since \(3\sim_\Gamma 1\), take \(x=1\), \(x_u=3\) and \(y=1\). One has
\begin{align}
    u_{11}u_{33}=\pi\pi' \neq 0,
    \qquad
    u_{11}u_{34}=\pi(\mathbb 1-\pi') \neq 0.
\end{align}
Thus there are two distinct values, \(y_a=3\) and \(y_a=4\), such that
\(u_{11}u_{3y_a}\neq 0\), so no definite value \(D_{11}(3)\) exists. Hence \(u\) is not differentiable.

To see explicitly that the action is not preserved, it is enough to choose a particularly simple \(\hat g_1\). Let
\begin{align}
    \hat g_1(1,3,3) = \hat G \neq 0,
\end{align}
and all other \(\hat g_1(x,x_u,x_v)\) vanish. Then in the \(O_1\) reference the action is simply
\begin{align}
    S
    =
    \Delta\hphi_1(1,3)\,\hat G\,\Delta\hphi_1(1,3).
\end{align}
Using the transformation rule for the field,
\begin{align}
    \hphi_1(x)=\sum_y u_{xy}\,\hphi_2(y)
    =\sum_y \hphi_2(y)\,u_{xy},
\end{align}
we have
\begin{align}\label{eq:transforem_phi_1_3_alt}
    \hphi_1(1)&=\pi\,\hphi_2(1)+(\mathbb 1-\pi)\,\hphi_2(2),
    \\\nonumber
    \hphi_1(3)&=\pi'\,\hphi_2(3)+(\mathbb 1-\pi')\,\hphi_2(4).
\end{align}
Hence
\begin{align}
    \| 3 - 1 \| \Delta \hphi_1(1,3)     =&
    \hphi_1(3)-\hphi_1(1)
    \\\nonumber
    =&
    \pi'\hphi_2(3)+(\mathbb 1-\pi')\hphi_2(4)
    \\\nonumber
    &-\pi\hphi_2(1)-(\mathbb 1-\pi)\hphi_2(2).
\end{align}
Substituting this, together with the corresponding expansion with the field operators to the left of the QP entries, gives
\begin{align}
    S = \frac{1}{4} \Big(
    &\hphi_2(3)\pi' + \hphi_2(4)(\mathbb 1-\pi')
    \\\nonumber
    &-\hphi_2(1)\pi - \hphi_2(2)(\mathbb 1-\pi) \Big)
    \\\nonumber
    &\hat G
    \\\nonumber
    \Big(
    &\pi'\hphi_2(3)+(\mathbb 1-\pi')\hphi_2(4)
    \\\nonumber
    &-\pi\hphi_2(1)-(\mathbb 1-\pi)\hphi_2(2) \Big).
\end{align}
This can be written as
\begin{align}
    S
    =
    \sum_{i,j=1}^4
    \hphi_2(i)\,\hat A_{ij}\,\hphi_2(j),
\end{align}
with for example
\begin{align}
    \hat A_{31}
        =
    -\frac{1}{4}\pi'\hat G\pi
    \neq
    \hat A_{31}^\dagger.
\end{align}
This cannot come from an action of the form
\begin{align}
    \sum_{y,y_a,y_b}
    \Delta\hphi_2(y,y_a)\,
    \hat g_2(y,y_a,y_b)\,
    \Delta\hphi_2(y,y_b),
\end{align}
with
\begin{align}
    \hat g_2(y,y_a,y_b)
    =
    \hat g_2(y,y_a,y_b)^\dagger
    =
    \hat g_2(y,y_b,y_a),
\end{align}
since expanding this form yields \(\hat A_{ij}\) that are a linear combination of different \(\hat g_2(y,y_a,y_b)\) with real coefficients \(\pm1\). Therefore, such a form necessarily yields  \(\hat A_{ij} = \hat A_{ij}^\dagger \) for all $i,j$.

\section{Proof that a quantum permutation is quantum controlled if and only if all of its entries commute}\label{ap:proof}

Let $u=(u_{xy})_{1\leq x,y \leq N}$ be a `classical' quantum permutation, i.e.\! all $u_{xy}$ commute with each other. The proof of Lemma 2.4 in \cite{weberQuantumPermutationMatrices2023} constructs for such a $u$ a quantum permutation $w^{(a)}$ and a unitary $W_a$ on $\Hc_O$ (which they denote as $\Hc$) such that $w^{(a+1)}_{xy}=W_{a+1} w^{(a)}_{xy} W_{a+1}^\dagger$ and $w^{(1)}_{xy}=W_{1} u_{xy} W_{1}^\dagger$ for all $x,y$ and $a=1,...,N$. Moreover, their construction ensures that for the last quantum permutation in this sequence, $w:=w^{(N)}$, its entries $w_{xy}$ are all diagonal and containing only $0$ and $1$.

Let us define now the unitary $W:=W_N W_{N-1}\cdots W_1$. It follows from the above that $w_{xy} = W u_{xy} W^\dagger$ for all $x,y$. Therefore,
\begin{align}
    u_{xy} &= W^\dagger w_{xy} W =  W^\dagger  \sum_{k=1}^d (w_{xy})_{kk} \ket{k}\bra{k} \;\; W \\\nonumber
    &= \sum_k (\sigma_k)_{xy} \; \pi_k = \left( \sum_k \sigma_k \otimes \pi_k \right)_{xy} ,
\end{align}
where we defined operators $\pi_k := W^\dagger \ket{k}\bra{k}W$ on $\Hc_O$ and $\sigma_k$ such that $(\sigma_k)_{xy} = (w_{xy})_{kk} $. From $w$ being a quantum permutation and from $w_{xy}$ being diagonal and containing only $0$ and $1$, it follows that $\sigma_k$ are permutation matrices on $N$ elements. Moreover, it is easy to see that $\pi_k$ are orthogonal projections on $\Hc_O$ that sum to the identity. Therefore, $u$ fulfills the definition of a QC permutation.

\end{document}